\documentclass[iop,apj]{emulateapj}
\usepackage{amsmath,amssymb,amstext,float}

\usepackage[citecolor=black,linkcolor=black]{hyperref}
\usepackage{color}

\usepackage[all]{hypcap} 
\usepackage{aas_macros}
\usepackage{natbib}
\shorttitle{Helicity Preference and Coronal Structure}
\shortauthors{Knizhnik et al.}
\newcommand{\beg}[1]{\begin{equation}\label{#1}}
\newcommand{\done}{\end{equation}}
\newcommand{\pd}[2]{\frac{\partial #1}{\partial #2}}

\newcommand{\vecB}{\textbf{B}}
\newcommand{\vecA}{\textbf{A}}

\newcommand{\vecv}{\textbf{v}}

\newcommand{\unit}[1]{\hat{\textbf{#1}}}

\newcommand{\curl}[1]{\nabla\times{#1}}
\newcommand{\divv}[1]{\nabla\cdot{#1}}
\numberwithin{equation}{section}
\begin{document}

\title{The Role of Magnetic Helicity in Structuring the Solar Corona}
\author{K. J. Knizhnik\altaffilmark{1,2}, S. K. Antiochos\altaffilmark{2}, and C. R. DeVore\altaffilmark{2}}
\altaffiltext{1}{Department of Physics and Astronomy, The Johns Hopkins University, Batimore MD 21287}
\altaffiltext{2}{Heliophysics Science Division, NASA Goddard Space Flight Center, Greenbelt MD 20771}
\begin{abstract}
 Two of the most widely observed and yet most puzzling features of the
 Sun's magnetic field are coronal loops that are smooth and laminar
 and prominences/filaments that are strongly sheared. These two
 features would seem to be quite unrelated in that the loops are near
 their minimum-energy current-free state, whereas filaments are
 regions of high magnetic stress and intense electric currents.  We
 argue that, in fact, these two features are inextricably linked in
 that both are due to a single process: the injection of magnetic
 helicity into the corona by photospheric motions and the subsequent
 evolution of this helicity by coronal reconnection.  In this paper,
 we present numerical simulations of the response of a
 \citet{Parker72} corona to photospheric driving motions that have
 varying degrees of helicity preference. We obtain four main
 conclusions: 1) in agreement with the helicity condensation model of
 \citet{Antiochos13}, the inverse cascade of helicity by magnetic
 reconnection results in the formation of prominences/filaments
 localized about polarity inversion lines (PILs); 2) this same process
 removes most structure from the rest of the corona, resulting in
 smooth and laminar coronal loops; 3) the amount of remnant tangling
 in coronal loops is inversely dependent on the net helicity injected by
 the driving motions; and 4) the structure of the solar corona depends
 only on the helicity preference of the driving motions and not on
 their detailed time dependence. We discuss the implications of our
 results for high-resolution observations of the corona.
\end{abstract}

\maketitle

\section{introduction}\label{sec:intro}

A well-known feature of the solar magnetic field is the observation of
filament channels at photospheric polarity inversion lines
(PILs). These magnetic structures, situated in the upper chromosphere and lower
corona, underlie and support the cool plasma that comprises
prominences and filaments \citep{Martin98,Gaizauskas00}. Filament
channels are regions of highly sheared magnetic field, containing large amounts
of free energy that ultimately is converted into kinetic and
thermal energy of the plasma, as well as nonthermal particle energies
when filament channels erupt and drive coronal mass ejections (CMEs). The
shear inherent in the filament channels is a form of magnetic
helicity, and filament channels are known as dextral if they have
negative helicity and sinistral if they have positive
helicity. Observations indicate that dextral (sinistral) filament
channels dominate in the northern (southern) hemisphere
\citep[e.g.][]{Martin92, Rust94b, Zirker97, Pevtsov03}. This
hemispheric helicity rule has also been observed in quiet-Sun magnetic
fields \citep{Pevtsov01b}, sigmoids \citep{Rust96}, active-region
magnetic fields \citep{Seehafer90}, coronal mass ejections (CMEs), and
sunspot whorls \citep{Pevtsov14}. The strength of the preference
ranges from about 55\% in active-region filaments \citep{Martin94}, to
over 80\% in quiescent filaments \citep{Pevtsov03} and does not seem
to change with solar cycle \citep{Hale27, Martin94,Hagino02}.\par

A second, seemingly unrelated, feature of the solar magnetic field is
the observation of loops in the closed-field corona that
appear to be near their minimum energy, current-free state. These
loops have been observed at high resolution in \emph{Transition Region
and Coronal Explorer (TRACE)} XUV and X-ray images, such as the one
in Figure \ref{fig:obs}, where they are seen to be very smooth and
laminar, with little to no tangling \citep{Schrijver99}. In other
words, there appears to be very little magnetic helicity associated
with the topology of these loops. The picture of the corona that
emerges, therefore, is one in which magnetic helicity manifests itself
at specific locations, namely above PILs, while leaving the rest of
the corona generally smooth and quasi-potential.\par

Since the corona has very high Lundquist number, its magnetic helicity
is believed to originate by injection from the photosphere, either by
flux emergence or footpoint motions after emergence, and to be lost to the
heliosphere by flux opening in CMEs and streamer blowouts.
\emph{Solar Dynamics Observatory (SDO)} measurements of helicity
injection into the coronal field indicate that shearing and twisting
motions by the photosphere dominate the flux emergence \citep{Liu12},
meaning that the helicity budget of the corona is primarily due to the
jostling of existing flux at the surface, rather than due to new flux
emerging into the corona. Observations of photospheric convection show
that these footpoint motions are highly complex \citep{Schmieder14},
with convective cells appearing randomly throughout the photosphere,
and occur over a broad range of scales, of order minutes for granules and
days for supergranules \citep{Hirzberger08}. From the standpoint of
helicity injection, however, the important flows are those that
twist the field. Compression of the field caused by converging flows
is not expected to impart any helicity into the field, so these
flows cannot be responsible for the shear observed in filament
channels. The flows that twist up the field, in contrast, do inject a
net helicity into the corona. Therefore, we conclude that it is
sufficient to model the helicity injection into the corona with simple
twisting motions as has been done by many authors
\citep[e.g.][]{WilmotSmith10, Rappazzo13}. Such flows have routinely
been observed in helioseismic measurements
\citep{Duvall00,Gizon03,Komm07,Seligman14}. Vortical flows on the
scale of granules \citep[e.g.][]{Bonet08, Bonet10, VD11, VD15} and
supergranules \citep{Brandt88,Attie09} have also been observed.\par

These considerations make it very challenging to understand the
simultaneous presence in the corona of both filament channels and
coronal loops. Magnetic stress is, apparently, injected
throughout the solar photosphere, yet is almost nowhere to be found in
the corona, except in filament channels.  \citet{Antiochos13}
presented a new model for the formation of filament channels, magnetic
helicity condensation, based on the well-known inverse cascade of
magnetic helicity in turbulent systems. In the helicity condensation
model, photospheric convection imparts helicity into the coronal field,
and this helicity is then transported throughout the corona by
magnetic reconnection, which is well-known to conserve helicity
\citep{Woltjer58,Taylor74,Taylor86,Berger84b}. 
Surface convection imparts the same sense
of twist to adjacent flux tubes, which are then able to undergo
component reconnection at their contact point. This component
reconnection produces a single flux tube with an axial flux equal to
the sum of the two original axial fluxes, but encircled by the same
twist field present on each of the two original flux tubes. In this
way the helicity, in the form of twist, inverse-cascades to larger and
larger scales. The PIL forms a natural boundary of the flux system, so
that when the twist reaches this boundary, it cannot proceed further,
since all of the flux has already reconnected. The end result of this
process is a mostly axial (untwisted) internal field, and a highly
sheared (twisted) field at the PIL, precisely what is observed as a
filament channel. At the same time, the untwisted internal field
corresponds to the laminar coronal loops. In this way, the helicity
condensation model provides a natural mechanism for the simultaneous
formation of both highly sheared filament channels and relatively
untwisted coronal loops. In this model, these two seemingly unrelated
features of the solar atmosphere are actually created by the same
process \citep{Antiochos13}.\par

The helicity condensation model was initially simulated by
\citet{Zhao15}, who injected magnetic helicity into a plane-parallel
Parker corona \citep{Parker72}. These authors found that photospheric
motions that inject the same helicity everywhere form filament
channels at the PIL. Furthermore, randomizing the photospheric motions
while keeping the same helicity injection rate did not qualitatively
affect the accumulation of twist flux at the PILs. \citet{Zhao15}  also
tested the effect of injecting helicity of opposite signs on adjacent
flux tubes, and found that their fields could not reconnect due to the
twist components being co-aligned.\par

In subsequent work \citep[][hereafter KAD15]{Knizhnik15}, we 
rigorously tested the helicity condensation model. We found that 
it not only qualitatively produced results consistent with the 
properties of filament channels, but that the inverse cascade of
magnetic helicity due to reconnection produces a twist flux at 
the PIL that agrees quantitatively with the predictions of the 
helicity condensation model. Based on this result, we estimated that
with the helicity preference observed on the Sun, filament channels
will form in about a day or so, in agreement with observations of filament
channel formation \citep{Martin98, Gaizauskas00}. We showed that
helicity condensation agreed both qualitatively and quantitatively
with observed properties of filament channels, and that the process
produced relatively untwisted coronal loops everywhere except at the
PIL. These results, however, were obtained for a $100\%$ helicity
rule, meaning that all of the helicity injected into the corona was of
the same sign. An obvious question to be raised is: what happens
if a fraction of the injected helicity has the opposite sign? Indeed,
this is a more realistic scenario since, as described above, the
corona has a hemispheric helicity preference, rather than a rule, so
that some helicity of the non-preferred sign is injected into the
corona at all times.\par

It is reasonable to expect that injecting helicity of the opposite sign
into the corona would simply slow down the helicity condensation process. 
However, this result is not as straightforward as it may seem. The 
simulations of \citet{Zhao15} demonstrated that adjacent flux tubes have
difficulty reconnecting if they are twisted in opposite senses. As 
a result, the twist was unable to inverse-cascade to larger scales,
as is required to form the sheared filament channels and smooth
coronal loops. Even if reconnection between the adjacent flux tubes is
eventually achieved as a result of some instability -- such as ideal
kinking -- driving the interaction, the twist-flux cancellation is 
expected to be far from perfect. Substantial residual twist could 
remain in what otherwise would have been smooth, untwisted coronal 
loops. Therefore, it is important to test whether sheared filament 
channels and smooth coronal loops form when both signs of magnetic 
helicity are injected into the corona.\par

In this paper, we investigate the effect of varying helicity preference 
on the structure of the closed-field corona. We performed helicity-conserving
numerical simulations that inject helicity into a plane-parallel
Parker corona, as we did in KAD15. Extending our previous work, the 
fraction of helicity of each sign that is injected into the corona 
was varied. We report on three cases: 1) $100\%$ of the
injected helicity is positive; 2) $75\%$/$25$\% of the injected
helicity is positive/negative; and 3) $50$\%/$50$\% of the injected
helicity is positive/negative. To make the simulations as realistic as
possible, we also randomized the pattern of helicity injection. We
compare the simulations with a fixed pattern of helicity injection to
those with a randomized pattern of helicity injection.\par

The paper is organized as follows. In \S \ref{sec:model} we discuss
the setup and initialization of our numerical simulations.
In \S \ref{sec:Hinjection} we describe how magnetic helicity 
is injected into the domain, and how the helicity preference is
employed. In \S \ref{sec:Results} we discuss the results of our
simulations, exploring the formation of filament channels and the
smoothness of coronal loops for various helicity preferences, and
compare the simulations with fixed and randomized patterns of
helicity injection. We discuss the implications for understanding
coronal magnetic structure in \S \ref{sec:implications}. \par

\section{Numerical Model}\label{sec:model}

 The numerical model used in this study was described previously in KAD15. We solve the equations of magnetohydrodynamics (MHD) using the Adaptively Refined Magnetohydrodynamics Solver \citep[ARMS; ][]{DeVore08} in three Cartesian dimensions. The equations have the form
\beg{cont}
\pd{\rho}{t}+\divv{\rho\vecv}=0,
\done
\beg{momentum}
\pd{\rho\vecv}{t} + \divv{\left( \rho\vecv\vecv \right)} = - \nabla P + \frac{1}{4\pi} \left( \curl{\vecB} \right) \times \vecB,
\done
\beg{energy}
\pd{T}{t} + \divv{\left( T\vecv \right)} = \left( 2 - \gamma \right) T \divv{\vecv},
\done
\beg{induction}
\pd{\vecB}{t} = \curl{ \left( \vecv \times \vecB \right)}.
\done
Here $\rho$ is mass density, $T$ is temperature, $P$ is thermal pressure, $\gamma$ is the ratio of specific heats, $\vecv$ is velocity, $\vecB$ is magnetic field, and $t$ is time. We close the equations via the ideal gas equation,
\beg{ideal}
P = \rho RT,
\done
where $R$ is the gas constant.\par

ARMS uses finite-volume representations of the variables to solve the system of equations. Its Flux Corrected Transport algorithms \citep{DeVore91} provide minimal, though finite, numerical dissipation, which allows reconnection to occur. As a result, to a very good approximation, ARMS conserves the magnetic helicity in the system.\par

We set up a model coronal field that is initially straight and uniform between two plates, as shown in Figure \ref{fig:init}. In this model, straight flux tubes represent coronal loops whose apex is located in the center of the domain and with the boundaries representing the photosphere. Our domain size is $[0,L_x] \times [-L_y,L_y] \times [-L_z,L_z]$, where $x$ is taken normal to the photosphere (the vertical direction) and we set $L_x=1$, and $L_y=L_z=1.75$. At all six sides, we use zero-gradient conditions, and the four side walls have open boundary conditions. Closed boundary conditions are employed at the top and bottom, where the magnetic field is line tied at the high-$\beta$ photosphere. The footpoints of the field lines do not move in response to magnetic forces, but do respond to imposed boundary flows to mimic driving at the plasma-dominated photosphere.\par

As in our previous work, we set the initial, uniform values in our dimensionless simulations to $\rho_0=1$, $T_0=1$, $P_0=0.05$, and $B_0=\sqrt{4\pi}$. These choices set the gas constant, $R=0.05$, the Alfv\'en speed, $c_{A0}=B_0 / \sqrt{4\pi\rho_0} = 1$, and the plasma beta, $\beta_0=8\pi P_0/B_0^2=0.1$. $\beta\ll1$ corresponds to a magnetically dominated plasma, which is generally true of the corona. The results discussed below will be given in simulation time, which is normalized to the time required for an Alfv\'en wave at unit speed ($c_{A0}=1$) to travel the distance separating the top and bottom plates ($L_x=1$).\par

As before, we set up our convective cells in a hexagonal pattern, with 84 cells on the top and bottom plates. Each individual cell has the same spatial and temporal profiles described in KAD15. The angular velocity of each cell is given by
\beg{angular}
\Omega(r,t)=
\begin{cases}
 -\Omega_0g(r)f(t) & \quad \text{if } r\le a_0 \\
0 & \quad \text{if } r > a_0 \\ 
\end{cases}
\done
where
\beg{gofr}
g(r)=\left(\frac{r}{a_0}\right)^4-\left(\frac{r}{a_0}\right)^8
\done
and
\beg{foft}
f(t)=\frac{1}{2}\left[1-\cos\left(\frac{2\pi t}{\tau}\right)\right]
\done
with $\Omega_0=7.5$ and $r$ the cylindrical radial coordinate with respect to the center of the convective cell. The flow is confined to a circle of radius $a_0=0.125$, and the magnitude of the flow is ramped up and back down over a period $\tau=3.35$. As demonstrated in KAD15, this velocity profile for the convective cells conserves the normal magnetic field distribution, $\vecB_x$, on the photospheric boundaries.\par

The simulation mesh for this study is specified, as in KAD15, such that we resolve very finely that part of the domain where these photospheric flows are imposed and the coronal magnetic field is influenced by the surface stresses. We use $4\times14\times14$ elemental blocks to span the simulation domain, each containing $8\times8\times8$ uniform, cubic grid cells. In the highly resolved portion of our simulation volume, which included the convective cells, the lanes between them, the untwisted region in the interior, and a buffer region around the outer perimeter of the pattern, we applied two additional levels of refinement, such that each rotation was covered by 32 grid points across its diameter. Closer to the side walls, the grid was allowed to coarsen by two levels, such that the grid spacing near the walls was a factor of four larger than in the interior. With this grid distribution, the ratio of the smallest grid spacing to the height of the box was about 0.001, resulting in a magnetic Reynolds number $R_m \sim 10^3$.

The key difference from our previous work, described in detail below, is that the sense of rotation of each cell, as well as the angular orientation of the entire pattern of cells, is changed randomly during the course of the simulations. The top plate mirrors the bottom plate at all times. These variations are meant to model more faithfully the random nature of the Sun's surface convection. Below, we describe how the relative fraction of cells rotating in the opposite sense affects the amount of magnetic helicity injected into our simulations.\par

\section{Magnetic Helicity Injection}\label{sec:Hinjection}
Magnetic helicity is a topological quantity describing linkages in the magnetic field, such as twist, shear, and writhe. In a volume $V$ bounded by a surface $S$, which need not be a magnetic flux surface, the relative magnetic helicity is given by \citep{Finn85}
\beg{Hdef}
H = \int_V{\Big(\vecA+\vecA_P\Big)\cdot\Big(\vecB-\vecB_P\Big) \;dV}.
\done
Here $\vecB=\nabla\times\vecA$ is the magnetic field in the volume $V$, generated by the vector potential $\vecA$, and $\vecB_P = \nabla \times \vecA_P$ is a current-free field ($\nabla\times\vecB_P = 0$) satisfying $\vecB_P\cdot\unit{n}|_S = \vecB\cdot\unit{n}|_S$. The rate of change of the helicity in Equation (\ref{Hdef}) is given in ideal MHD by 
\beg{dHdtdef}
\frac{dH}{dt} = 2 \oint_S{\Big[\Big(\vecA_P\cdot\vecv\Big)\vecB - \Big(\vecA_P\cdot\vecB\Big)\vecv\Big]\cdot d\textbf{S}}.
\done
The first term represents the effects of motions on the boundary with velocity $\vecv$, while the second term represents the emergence or submergence of magnetic field through the boundary. As a result, the magnetic helicity in our simulation changes only due to motions on or through the boundary. In highly conducting ($R_m \gg 1$) plasmas, such as the corona, magnetic helicity is conserved even in the presence of a small localized resistivity that enables magnetic reconnection \citep{Woltjer58, Taylor74, Taylor86, Berger84b}. Since no new flux is being injected at our top and bottom boundaries, and the side boundaries are sufficiently far away that no flux leaves the system, the rate of change of magnetic helicity in our simulation reduces to
\beg{dHdtsim}
\frac{dH}{dt} = 2 \oint_S{\Big(\vecA_P\cdot\vecv\Big)\vecB\cdot d\textbf{S}}.
\done
The magnetic helicity $H_0$ injected into a single flux tube (i.e. one top/bottom pair of rotation cells) over one cycle is obtained by integrating Equation \ref{dHdtsim} from $t=0$ to $t=\tau$, employing Equations \ref{angular}--\ref{foft} and the vector potential for the uniform, current-free initial field,
\beg{Ap}
\vecA_P = \frac{B_0}{2}\Big(y\unit{z}-z\unit{y}\Big).
\done

If the sense of rotation is clockwise, the resulting positive helicity injected is 
\beg{helicity1cell}
H_0 = 2\times10^{-2}.
\done
Trivially, the net helicity $H_\Sigma$ injected into $N$ such flux tubes all twisted in the clockwise sense, as in KAD15, is 
\beg{NH1KAD15}
H_\Sigma = N H_0.
\done
In this paper, we generalize to cases in which $N_+$/$N_-$ cells rotate clockwise/counter-clockwise and inject positive/negative helicity, with $N_+ + N_- = N$. The net helicity injected into the corona then becomes 
\beg{helicityallN}
H_\Sigma = \left( N_+ - N_- \right) H_0.
\done
The case studied in KAD15 has $N_+ = N$ and $N_- = 0$, so Equation (\ref{helicityallN}) reduces to Equation (\ref{NH1KAD15}). If, on the other hand, $N_+ = N_-$, equal numbers of cells rotate in each sense and the net injected helicity vanishes, $H_\Sigma = 0$. In our simulations described below, we allowed $N_+$ and $N_-$ to vary from cycle to cycle. The net helicity injected into the corona after $M$ cycles therefore is 
\beg{Htotal}
H_\Sigma = \sum_{i=1}^{M} \left( N_{+,i} - N_{-,i} \right) H_0,
\done
where $N_{+,i}/N_{-,i}$ is the number of cells that rotate clockwise/counter-clockwise during cycle $i$.\par

In each simulation, we assign a probability $k$ that any individual top/bottom pair of cells injects positive helicity and $1-k$ that the pair injects negative helicity. A random number $\kappa_j \in [0,1]$ is generated for each pair of cells $j \in [1,N]$ during each cycle, and $\kappa_j$ is compared with $k$ to determine whether the sense of rotation is clockwise or counter-clockwise over that cycle. The helicity $H_j$ injected by the $j$th pair of cells is 
\beg{Hj}
H_j = H_0 \times 
\begin{cases}
+1 & \text{if } \kappa_j \le k; \\
-1 & \text{if } \kappa_j > k. \\
\end{cases}
\done
On average, the expectation is that during each cycle, a fraction $2k-1$ of the maximum positive helicity $N H_0$ will be injected into the corona, 
\beg{Hk}
\langle H_\Sigma \rangle = (2k-1) N H_0 = f N H_0.
\done
Throughout the paper, we will refer to $k$ as the {\rm helicity preference} of each simulation, and to $f = 2k-1$ as the {\rm net fractional helicity} associated with $k$.\par

For this paper, the cases $k=0.75$ and $k=0.5$ were simulated to complement the $k=1$ case previously presented in KAD15. For reference, the expectation values of the net helicity injected per cycle are 
\beg{helicityk}
\langle H_\Sigma \rangle = f N H_0 = N H_0 \times
\begin{cases}
1.0 & \text{if } k=1.0; \\ 
0.5 & \text{if } k=0.75; \\
0.0 & \text{if } k=0.5. \\
\end{cases}
\done
Below, we use the precise number of cells injecting each sign of helicity during each cycle to evaluate $H_\Sigma$ in Equation (\ref{Htotal}). That prediction is compared to the instantaneous value $H(t)$ calculated directly from the \citet{Finn85} volume integral for the relative magnetic helicity in the simulation, as described in KAD15.\par

The helicity preference $k$ introduces one aspect of randomness into our simulations through the assignment of a clockwise/counter-clockwise sense of rotation (and positive/negative helicity) to each pair of rotation cells during each cycle of rotation. In order to emulate the stochastically shifting spatial pattern of convection on the solar surface and investigate its effect on coronal structure, we also introduce a second aspect of randomness into a separate set of simulations. After each cycle of rotations, we displace the entire hexagonal cellular pattern shown in Figure \ref{fig:init} by a randomly chosen angle $\theta \in [0^\circ,60^\circ]$ about its central vertical axis $(y,z) = (0,0)$. For simplicity, the same angular displacement is applied to both the top and bottom plates, so that the top/bottom pairs of rotation cells remain aligned as before. Now, however, the random displacement means that the rotation cells will, in general, encompass parts of multiple neighboring flux tubes that were twisted during the previous cycle of rotations. The ensuing cycle therefore introduces braiding, as well as twisting, into the coronal magnetic field between the plates. As we will show, however, this displacement-induced braiding has no effect on the rate of helicity accumulation in the corona, and has only a minor influence on the smoothness of the induced magnetic structure. This result concurs with the qualitative conclusions of \citet{Zhao15} from a much simpler simulation setup and is analyzed quantitatively here for the first time.\par

\section{Results}\label{sec:Results}

In this section, we first describe the results of our simulations with both fixed and randomized patterns and for the various helicity preferences. Then, we analyze those results in the context of filament-channel formation and the smoothness of coronal loops.\par

\subsection{Fixed Cellular Pattern}

The first set of simulations holds the cellular pattern fixed in the orientation shown in Figure \ref{fig:init}, randomizing only the sense of rotation of the individual cells as described in \S \ref{sec:model}. To compare the $k=0.75$ simulation most consistently with the $k=1.0$ case presented in KAD15, we ran it for twice as many cycles (42 vs.\ 21). As shown by Equation (\ref{Htotal}), the expectation value for the net injected helicity is the same (50\% of $42 \times N$ vs.\ 100\% of $21 \times N$). The $k=0.5$ case, in contrast, accumulates a net helicity only due to statistical fluctuations away from its average value of zero. For that case, therefore, we simply ran the simulation for $21$ cycles. All three simulations then were extended for $5$ additional cycles without imposing any rotational motions, to allow transients to die down and the system to relax toward a quasi-equilibrium final state.\par

As an example, Figure \ref{fig:vphi_f} shows the azimuthal component of the velocity on the bottom plate, $V_\phi(x=0,y,z)$, during the first (left) and second (right) cycles of the $k=0.75$ case. For both the $k=0.75$ and $k=0.5$ cases, each individual cell does not necessarily preserve its sense of rotation from one cycle to the next. The sense of rotation is assigned randomly at each cycle, as given above in Equation (\ref{Hj}). For $k=1.0$, all cells rotate in the same sense in the first and second -- indeed, throughout all -- cycles. \par

Figure \ref{fig:helicityf} shows the analytically expected (solid line) and numerically calculated (dashed line) helicities for each simulation. The orange ($k=1.0$) curve is the same as that presented in KAD15. After $21$ cycles, the rotation cells have injected $H=36$ units of helicity. The red ($k=0.75$) curve shows that approximately the same $H=36$ units are injected over twice the time (cf.\ Equation \ref{helicityk}) for the 75\% preference. The blue ($k=0.5$) curve shows, as expected, that almost no net helicity is accumulated in the simulation with 50\% preference over its first $21$ cycles. All of the numerically calculated curves match very well the analytical values at each cycle, demonstrating that our simulations conserve helicity to a very high degree of accuracy. Therefore, the evolution of the magnetic field in our simulations is due predominantly to convection and reconnection, rather than to numerical diffusion that would dissipate helicity.\par

\subsection{Randomly Displaced Patterns}

The second set of simulations is identical to the first, except that we also displace the entire hexagonal cellular pattern through a random angle after each cycle as described in \S \ref{sec:model}. In these setups, different flux tubes wrap around each other, creating a braided field, in addition to being twisted by the rotation cells. We ran each simulation for the same number of cycles as in the fixed-pattern cases and for the same values of $k$.\par

As an example, Figure \ref{fig:vphi_r} shows $V_\phi(x=0,y,z)$ during the first (left) and second (right) cycles of the $k=0.75$ case. Like the corresponding fixed pattern cases, in the random $k=0.75$ and $k=0.5$ cases, each individual cell does not necessarily maintain its sense of rotation. Unlike the fixed pattern cases, however, the pattern itself is displaced by a random angle after each cycle. Except for the $k=1.0$ case, the randomized patterns (Fig.\ \ref{fig:vphi_r}) exhibit different distributions of color than the fixed patterns (Fig.\ \ref{fig:vphi_f}). We used different sequences of random numbers $\kappa_j$ to set the clockwise/counter-clockwise sense of rotation of the individual rotation cells in the two sets of simulations. The random angular displacements of the cellular pattern between the first and second cycles are evident by comparing the left and right columns for each helicity preference.\par

Figure \ref{fig:helicityr} shows the analytically expected (solid line) and numerically calculated (dashed line) helicities for the various cases. Although the average helicities (Eq.\ \ref{helicityk}) injected into the corona are identical for each value of $k$, the precise helicities (Eq.\ \ref{Htotal}) actually injected differ between the fixed and randomized cases due to statistical fluctuations. Thus, the curves in Figures \ref{fig:helicityf} and \ref{fig:helicityr} are slightly different for $k \ne 1.0$. The orange ($k=1.0$) curve shows that the helicity injected for the 100\% preference is identical for the fixed and randomized patterns, as expected. The red ($k=0.75$) curve shows that the 75\% preference injects slightly more helicity in twice the time. The blue ($k=0.50$) curve shows that, as before, almost no net helicity is injected in this case. In this randomized-pattern simulation, the residual net helicity for the 50\% preference is small and positive, whereas in the fixed-pattern simulation, it is negative. In all cases, we again find excellent agreement between the numerically calculated and analytically expected helicities.\par


\subsection{Formation of Filament Channels}

Figure \ref{fig:Bphi_beg} shows the azimuthal component of the magnetic field in the horizontal mid-plane, $B_\phi(x=0.5,y,z)$, halfway through the first cycle of twist for each simulation. At this early stage, each case exhibits the characteristic hexagonal pattern of rotation cells. For $k=1.0$, every cell injects the same sign of $B_\phi$, so adjacent flux tubes always have oppositely directed twist fields and are able to reconnect readily. For the $k=0.75$ and $k=0.5$ cases, in contrast, adjacent twist fields sometimes are in the same direction. On average, this is true half the time in the $k=0.5$ case, suppressing reconnection between adjacent flux tubes whose twist fields are parallel rather than anti-parallel.\par

The effect of the helicity preference on the formation of filament channels can be seen clearly in the final-time $B_\phi$ maps in Figure \ref{fig:Bphi_end}. The $k=1.0$ case has accumulated oppositely signed bands of twist flux at the outer and inner boundaries of the hexagonal pattern, as described previously in KAD15. These bands result from the inverse-cascade of twist flux from small to large scales due to reconnection, collecting at the boundaries to form filament channels according to the helicity-condensation model \citep{Antiochos13}. The $k=0.75$ case has been run out twice as long in order to accumulate roughly the same helicity, and it has acquired similar bands of twist at the outer and inner boundaries of the hexagonal pattern. Thus, despite the one-third (25\%/75\%) of twist fields on neighboring flux tubes that are parallel rather than anti-parallel in this simulation, sufficient reconnection has occurred to enable the helicity to condense at the flux-system boundaries here, as well. The shapes of the filament channels differ slightly in the $k=1.0$ and $k=0.75$ cases. The contrast is most evident in the randomized-pattern simulations, where the twist flux has a very uniform, circular appearance in the $k=1.0$ case, while the structure is more ragged, especially at the inner boundary, in the $k=0.75$ case. At the largest scales, however, these two cases yield qualitatively identical outcomes: the twist flux forms two bands of opposite polarity at the boundaries of the hexagonal pattern of rotations. In the corona, such bands would manifest themselves as extended, sheared filament channels.\par

The sharply contrasting $k=0.5$ case, on the other hand, displays a very different final-time appearance. No long, coherent bands of twist flux have accumulated at either the outer or inner boundaries of the hexagonal pattern. Instead, there are localized concentrations of twist flux dispersed across the interior of the pattern, as well as at its boundaries. Because zero net helicity is injected into this system, on average, zero net twist flux is available to be transported by reconnection to the hexagonal boundaries where it can accumulate. Turning this argument around, if the net condensed twist flux were finite, then the net helicity would be finite as well. We demonstrated this result analytically in KAD15. Consequently, the helicity-condensation process does not form filament channels in the case of a 50\% helicity preference.\par

These examples demonstrate that the helicity preference plays a major role in the organization of the twist flux and the formation of filament channels. The $k=0.75$ case forms similarly strong, although rather more structured, bands of twist flux over twice the time as the $k=1.0$ case. As is argued below, the time scale for filament-channel formation is inversely proportional to the average net fractional helicity injected, i.e.\ to $f$. This dependence is supported further by the absence of filament-channel structure in the $k=0.5$ case, whose predicted time scale for channel formation is infinite.\par

\subsection{Accumulation of Twist Flux}\label{sec:twistflux}

The results above demonstrate that there are major qualitative and quantitative contrasts between the results for different helicity preferences, but more minor differences between the fixed and random patterns for a given helicity preference. We begin the quantitative analysis of our simulations by calculating the positive twist flux $\Phi_{tw}^+$ through the $z=0$ plane, 
\beg{Phitw}
\Phi_{tw}^+ = \int_0^{L_x}{dx} \int_0^{L_y}{dy \; B_{tw}^+(x,y,z=0)},
\done
where the corresponding positive twist field $B_{tw}^+$ is 
\beg{Btw}
B_{tw}^+ = \frac{1}{2} \left( B_\phi + \left\vert B_\phi \right\vert \right) \ge 0.
\done
The twist flux $\Phi_{tw}^+$ is plotted in Figure \ref{fig:tw_v_time} for both the fixed (solid curves) and random (dashed curves) patterns. All six simulations exhibit a brief initial phase of ideal evolution, of about one rotation cycle in duration, in which twist flux is injected into and stored in individual, non-interacting flux tubes. At this stage, the sense of rotation of adjacent cells is irrelevant to the accumulation of twist flux. As the twisting continues, however, the flux tubes expand laterally to compress the volume between them. This forms and strengthens electric current sheets between neighboring tubes that have anti-parallel twist fields. Reconnection between such tubes commences during subsequent twist cycles. This process, together with the randomization of the sense of rotation of individual cells (for $k \ne 1.0$) and of the orientation of the cellular pattern (for the random cases), causes the curves to deviate increasingly from one another at later times.

The two cases with nonzero net fractional helicities, $k=1.0$ (orange) and $k=0.75$ (red), show relatively small differences between the fixed and random patterns for fixed $k$. Over the full duration of the simulations, each preference accumulates essentially the same twist flux. The slightly larger values for the $k = 0.75$ case reflect the slightly larger magnetic helicities accumulated in those simulations (Figs.\ \ref{fig:helicityf} and \ref{fig:helicityr}) compared to the $k = 1.0$ case. All four of these simulations eventually accumulate twist flux at a rate per cycle that is in good agreement with the calculation by KAD15 (their Equation 4.13 and Figure 11),
\beg{DPhitw}
\Delta \Phi_{tw} = \frac{1}{2} \frac{\Delta \langle H_\Sigma \rangle}{\Phi_N} = \frac{f}{2} \frac{H_0}{\Phi_0},
\done
where each of the $N$ twisted flux tubes contains $\Phi_0$ units of magnetic flux. The expression in Equation (\ref{DPhitw}) assumes that the twist flux $\Phi_{tw}$ condenses at the outer boundary of the flux system, which occurs in our simulations with $k = 1.0$ and $k = 0.75$. In a spirit similar to KAD15 (their Equation 4.28), we calculate the filament-channel formation time $\tau_{fc}$ over which a critical amount of twist magnetic flux $\Delta \Phi_{fc}$ accumulates, 
\beg{Taufc}
\tau_{fc} = \frac{\Delta \Phi_{fc}}{\Delta \Phi_{tw}} \tau_0 = \frac{2}{f} \left( \frac{\Phi_0}{H_0} \tau_0 \right) \Delta \Phi_{fc},
\done
where $\tau_0$ is the duration of one twist cycle. Equations (\ref{DPhitw}) and (\ref{Taufc}) quantitatively express the observed factor-of-two differences in twist-flux accumulation rates and filament-channel formation times between our $k = 1.0$ ($f = 1.0$) and $k = 0.75$ ($f = 0.5$) cases. They also predict how these quantities should change for other helicity preferences $k$.\par

Our last two simulations, with $k=0.5$, have zero net fractional helicity, $f = 0.0$. For this case, the predicted accumulated twist flux $\Delta \Phi_{tw}$ vanishes and the filament-channel formation time $\tau_{fc}$ is infinite. We observe in these simulations (Figure \ref{fig:tw_v_time}) that the fluctuations in the twist flux are relatively large, and the average amount of flux saturates after about 10 cycles have elapsed. Thereafter, the average seems to be statistically quasi-steady, increasing or decreasing randomly according to the cycle-to-cycle variations of the sign of twist in individual rotation cells (in both simulations) and of the orientation of the cellular pattern (in the random-pattern simulation only). Evidently, these simulations have reached a roughly steady-state balance between the rates of twist-flux injection by the twisting motions and extraction via a combination of untwisting motions and reconnection between anti-parallel twist fields.\par

\subsection{Smoothness of Coronal Loops}

We have seen that, when it is effective, the helicity-condensation
process transports twist via reconnection to the boundaries of the
flux system, where it condenses. This leaves the interior of the
system relatively smooth and untwisted. The final configuration then
corresponds to a corona with strong shear concentrated at its PILs and
laminar coronal loops in interior regions away from its PILs. This result can
be seen clearly in Figure \ref{fig:Bphi_end}. In the $k=1.0$ and $k=0.75$
cases with nonzero net fractional helicities, at a glance, the interior
of each flux system seems very smooth, with little twist evident. A
careful comparison of the two cases reveals that the annular region
between the filament channels is somewhat more structured for
$k=0.75$, with localized, small-amplitude twists of both signs
accumulating in the interior. As might be anticipated, this structure
is somewhat less noticeable for the simulations with randomly
displaced patterns compared to their fixed-pattern counterparts.\par

The appearance of the $k=0.5$
is strikingly different from the $k=1.0$ and $k=0.75$
cases. For the fixed pattern especially (bottom row, left column of Figure
\ref{fig:Bphi_end}), small-scale, coherent concentrations
of twisted field are present throughout the interior of the hexagonal
flow region. In addition, the magnitude of the accumulated twist is
significantly smaller than in the $k = 1.0$ and $k = 0.75$ cases. This
is due to both the random untwisting of previously twisted field lines
in successive cycles and the zero net twist flux that can accumulate
globally and be transported to the flux-system boundaries. The local
twist concentrations that are formed appear and disappear transiently
as the system evolves. Each such concentration has a lifetime on the
order of one rotation period of the convection cells. 
  Taking the rotation period to be of order a day, or the lifetime of
  a typical supergranule, these concentrations of twist should easily
  survive for timescales long enough to be detected remotely. The lack
  of such observations indicates that the photosphere likely injects
  helicity with a significant preference.\par

 To demonstrate the stark difference in the amount of structure in the different helicity preference cases we plot in Fig. \ref{fig:structure_f} a set of magnetic field lines from the same set of fixed points for the fixed-pattern $k=1$, $k=0.75$, and $k=0.5$ cases. All of the field lines are chosen from the interior of the hexagonal region, which represents the `loop' portion of the corona. The bottom plate represents magnetic field magnitude, which shows a structure similar to that shown in Fig. \ref{fig:Bphi_end}. Although only a sample of field lines is chosen, they are representative of the field lines in the rest of the `loop' portion of the corona. The striking difference in the amount of structure in the corona is immediately evident by comparing the $k=1$ and $k=0.75$ cases with the $k=0.5$ case. In the latter simulation, field lines are twisted and braided around each other in a complicated fashion. In the first two cases, although the field lines are traced from the same points, the field lines themselves are quite smooth and laminar. There may appear to be some structure due to field lines passing behind, or in front of, each other, but there is almost no significant twisting or braiding around each other. In fact, the top two figures very closely resemble the initial, uniform field configuration, and, importantly, the smooth, laminar structure observed in Fig. \ref{fig:obs}, albeit in a plane-parallel geometry. In this sense, the $k=1$ and $k=0.75$ coronal loops are quasi-potential, while the $k=0.5$ coronal loops clearly deviate quite strongly from quasi-potentiality. In Fig. \ref{fig:structure_r}, a different set of field lines is plotted for the random-pattern $k=1$, $k=0.75$ and $k=0.5$ cases. These sample field lines are traced from the same footpoints in each case, and again demonstrate the quasi-potentiality of the corona in the former two cases, and the large amount of structure in the latter case. \par

\subsection{Fluctuations of Twist Field}

To quantify the amount of small-scale structure in the various cases, we calculated the angle-averaged azimuthal magnetic field $\langle B_\phi(r) \rangle$ and its root mean square deviation $\delta B_\phi(r)$ in the mid plane $(x = 0.5)$. Specifically, we evaluated (for $m = 1,2$) 
\begin{align}
\label{Bm}
\langle B_\phi^m(r) \rangle &= \frac{1}{2\pi} \int_0^{2\pi} B_\phi^m(x=0.5,y,z) d\phi,\\
\label{dB}
\delta B_\phi(r) &= \sqrt{\langle B_\phi^2(r) \rangle - \langle B_\phi(r) \rangle^2},
\end{align}
where $r = \sqrt{y^2+z^2}$. A discrete $\rho$ grid, with the same
spacing as the $y$ and $z$ grids, was adopted, and all cell-center
positions $(y,z)$ were grouped into corresponding $r$ intervals to
calculate the integrals in Equation (\ref{Bm}).
  Figure \ref{fig:width} shows $\delta B_\phi(r)$ at the end of the
  random pattern simulations for each value of $k$ (color-coded). The
  rms deviations $\delta B_\phi(r)$ shown in Figure \ref{fig:width}
  all exhibit small-scale statistical fluctuations. The average
  amplitude of the fluctuations is smallest for the 100\% helicity
  preference ($k = 1.0$, orange) and largest for the 50\% preference
  ($k = 0.5$, blue), where they are nearly three times as large
  compared to the 100\% helicity preference. A very similar trend was
  seen in the fixed-pattern simulations. As evident from Figure 11,
these localized fluctuations in twist field for the zero helicity
case correspond to field line tangling that would easily be
observed if it were present in the real corona.\par

\subsection{Length of Field Lines}

Further insight into the overall magnetic structure of the corona is gleaned by examining the lengths of field lines throughout the domain. In Figure \ref{fig:Blength}, we plot the length of magnetic field lines for the various helicity preferences. Here the differences between the filament channels and coronal loops is greatest in the $k = 1.0$ simulations, where substantially longer field lines reside in the filament channels. The coronal loops in the interior are quite short. Indeed, their lengths are very close to those of the untwisted field lines exterior to the hexagonal region. The two $k = 0.75$ simulations, meanwhile, display a slightly more mixed character in the interior. The coronal-loop field lines in these cases are slightly longer than the exterior untwisted field, although not nearly as long as the filament-channel field lines, which themselves are somewhat shorter than for $k = 1.0$.\par

As we found for several diagnostics described previously, the $k = 0.5$ case displays a strikingly different appearance from those with nonzero net fractional helicities $f$. As can be seen in Figure \ref{fig:Blength} (bottom row), the hexagonal region of rotation cells hosts a rather homogeneous mixture of relatively short field lines, structured at small scales. This is not dissimilar to the interior of the hexagonal region in the $k = 0.75$ case (middle row). However, no large-scale organization of the field-line length is evident, beyond the exclusion of twist from the center of the domain and the region beyond the perimeter of the hexagonal region. This is in sharp contrast to the cases $k = 1.0$ and $0.75$, where much longer field lines accrue at both the inner and outer boundaries of the flux system.\par
\par

\section{Implications for Coronal Structure}\label{sec:implications}

The results described in the preceding section have important implications for the global structure of the solar corona. Our findings demonstrate that the magnetic helicity preference $k$ plays key roles in determining how the corona is structured and the time scale over which that structure develops. The contrast is particularly strong between the cases with 100\% ($k = 1.0$) and 50\% ($k = 0.5$) preferences and randomly displaced cellular patterns shown at the top and bottom right, respectively, in Figures \ref{fig:Bphi_end} and \ref{fig:Blength}. For the 100\% preference, the twist flux condenses into two primary bands with opposite senses of twist at the inner and outer boundaries of the hexagonal region of rotation cells, with very little twist in the interior. These concentrations and dilution are reflected in the lengths of the associated magnetic field lines, which are very long near the two boundaries but minimally short in the interior. For the 50\% preference, on the other hand, the twist flux does not condense into any recognizable global-scale structure, and the field lines have an essentially homogeneous distribution of intermediate lengths. These two cases have the largest and smallest (in magnitude) net fractional helicities, $f = 1.0$ and $f = 0.0$, respectively.\par

Our intermediate case with 75\% preference ($k = 0.75$, $f = 0.5$) exhibits some features of both of the previous limiting cases but, importantly, qualitatively resembles more closely the results for the 100\% preference. The bands of condensed twist flux still form, albeit twice as slowly and with significant intrusions of twisted structures between them, and the field-line lengths are correspondingly longer at the hexagonal-region boundaries than in its interior, although with less contrast. Extrapolating to other cases with even smaller preferences but nonzero fractional helicities -- say, $k = 0.625$ and $f = 0.25$ -- we would expect these trends to continue, with a further increase in the filament-channel formation time (another doubling for $f = 0.25$) and in the amount and homogeneity of small-scale structure in the interior of the hexagonal region.\par

\begin{figure*}[!p]
\centering\includegraphics[scale=0.65,trim=0.0cm 0.0cm 0.0cm 0.0cm,clip=true]{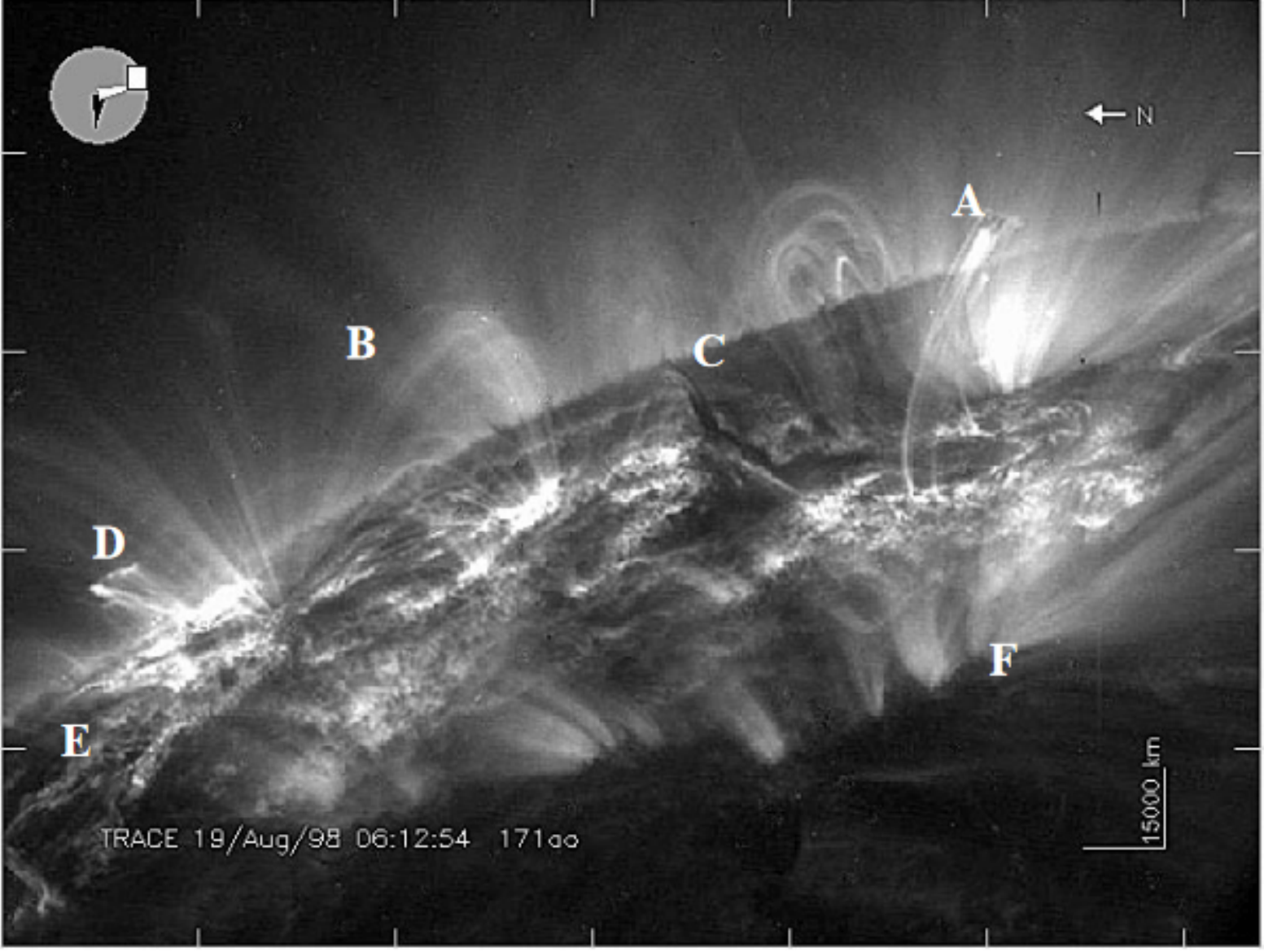}
\caption{Coronal loops as observed by TRACE, showing their quasi-potential structure, with little to no tangling. \citep[From][]{Schrijver99}}
\label{fig:obs}
\end{figure*}

\begin{figure*}[!p]
\centering\includegraphics[scale=0.65,trim=0.0cm 0.0cm 0.0cm 0.0cm,clip=true]{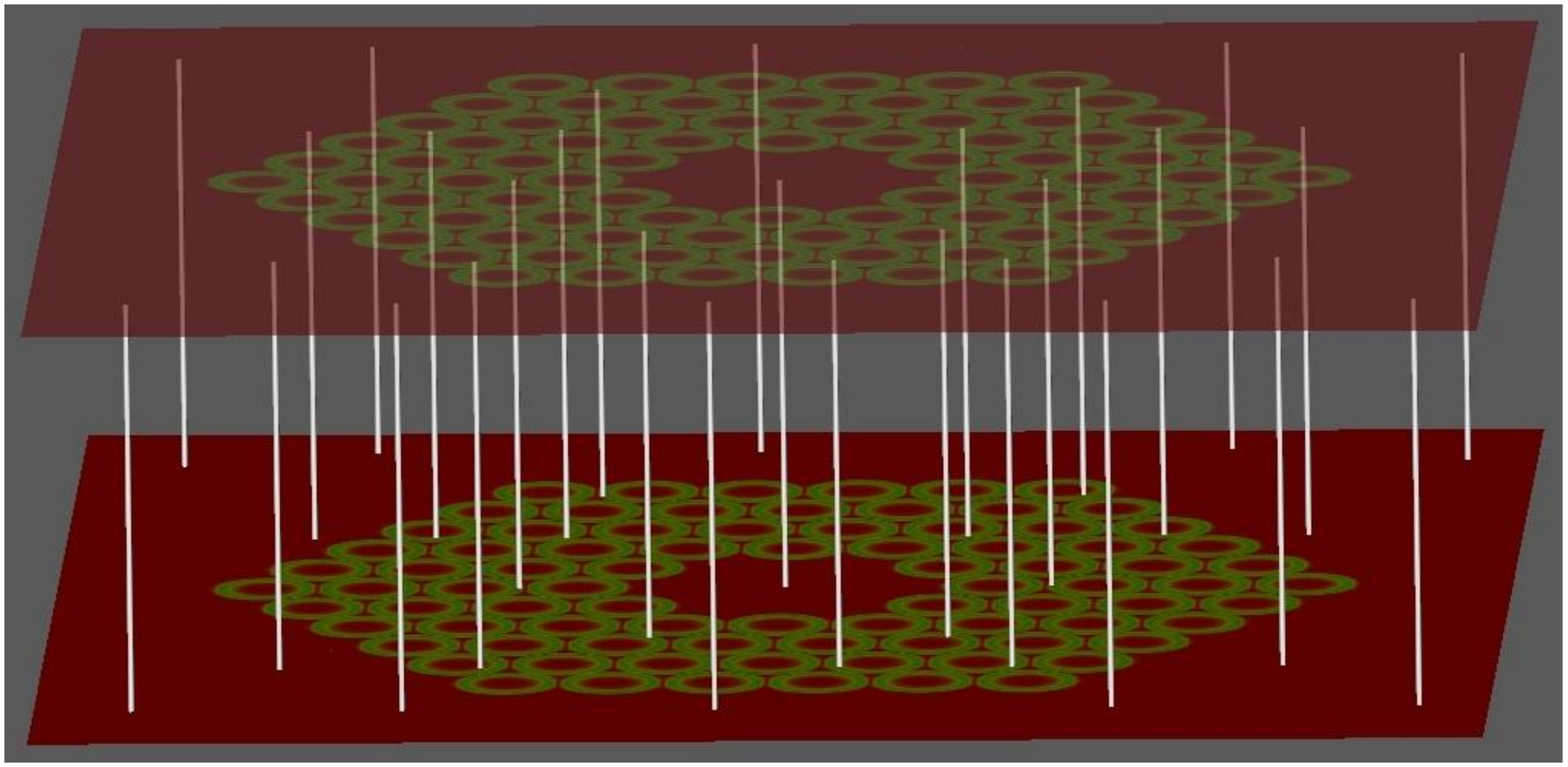}
\caption{Setup of the numerical simulations. White lines represent the initial vertical magnetic field; color shading on the to and bottom plates represents velocity magnitude. To emulate the random photospheric convection in our numerical experiments, the local sense of rotation (clockwise/counter-clockwise) of individual convective cells shown in this figure can be set randomly, and the global hexagonal pattern of the cells collectively can be rotated randomly about its center. See text for details.}
\label{fig:init}
\end{figure*}

\begin{figure*}
\centering\includegraphics[scale=0.4,trim=0cm 0cm 3.5cm 0cm, clip=true]{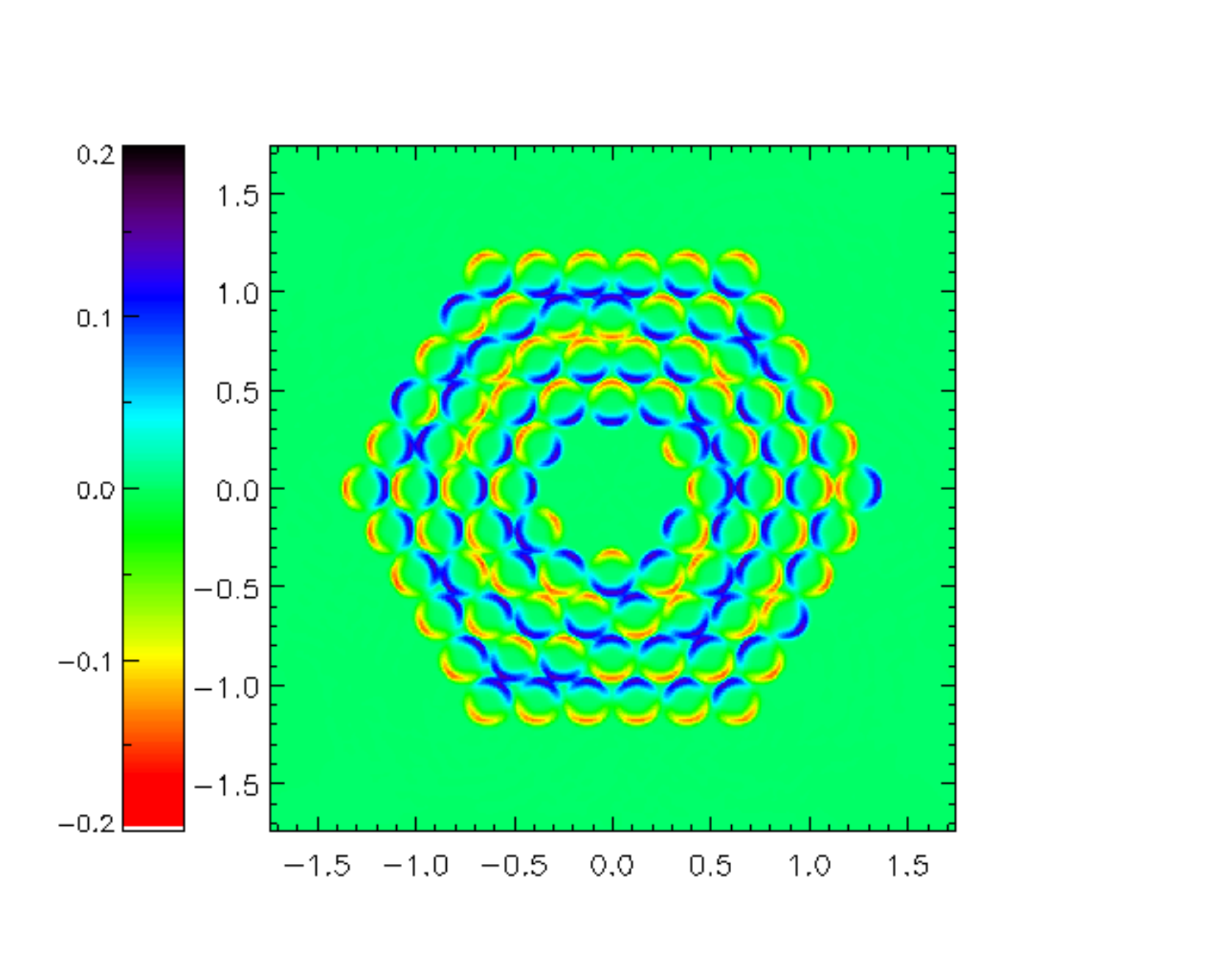}
\centering\includegraphics[scale=0.4,trim=0cm 0cm 3.5cm 0cm, clip=true]{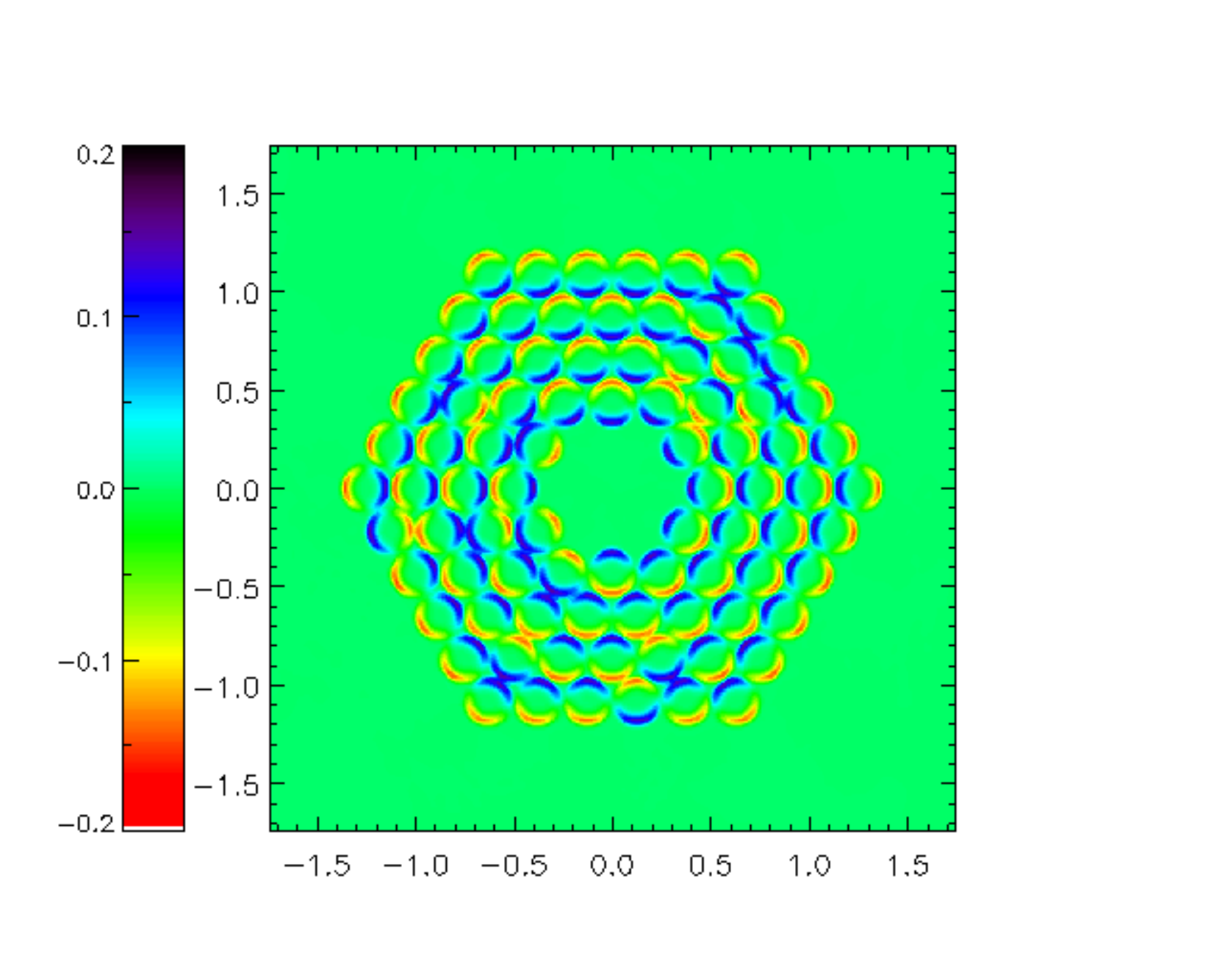}
\caption{$V_\phi$ on the bottom plate for the fixed-pattern case with $k=0.75$ during the first (left) and second (right) cycles.}
\label{fig:vphi_f}
\end{figure*}

\begin{figure*}[!p]
\centering\includegraphics[scale=0.75,trim=0.0cm 6.5cm 0.0cm 0.0cm]{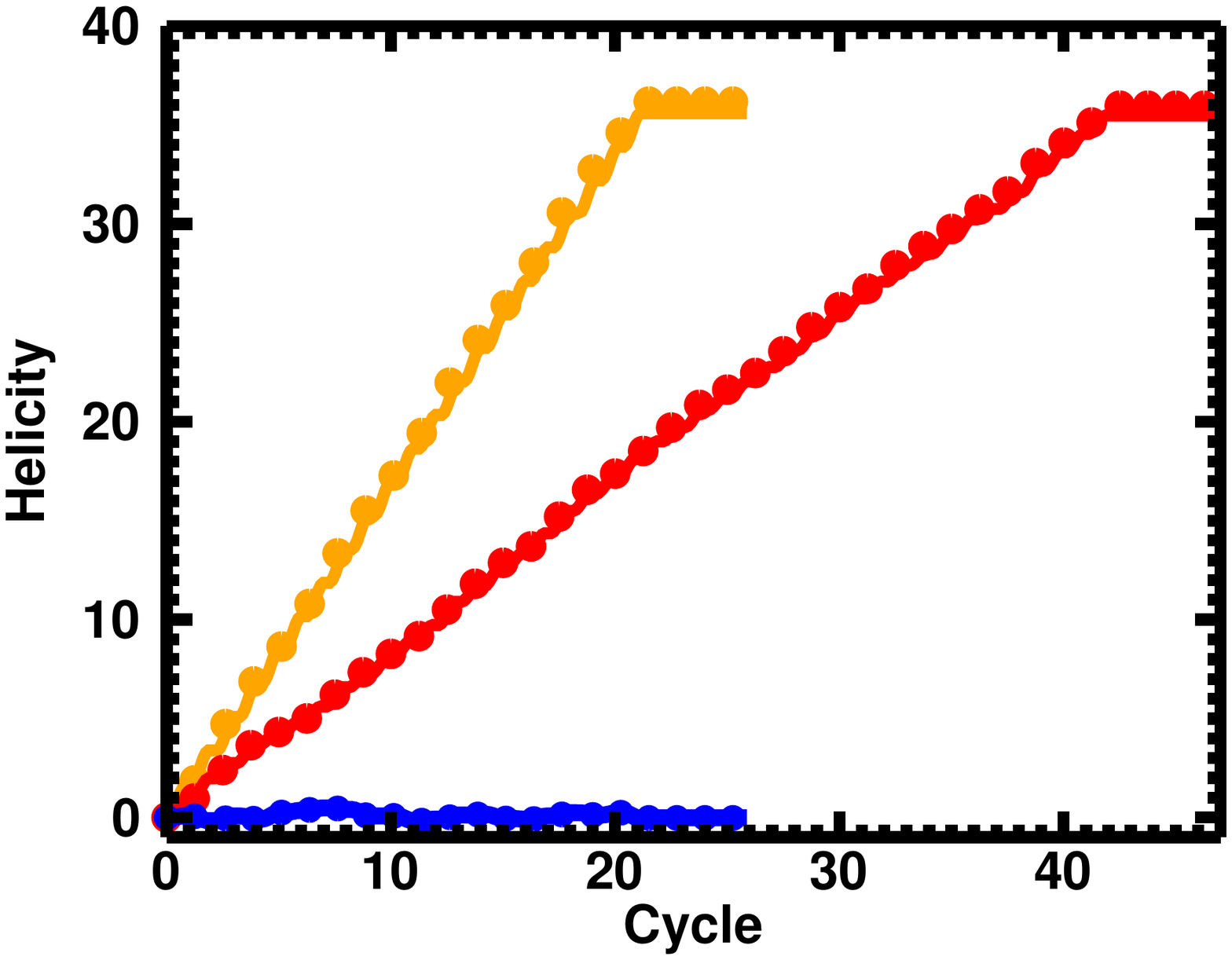}
\caption{Helicity for fixed-pattern simulations. Solid curves are the analytically expected helicity based on the number of convective cells injecting positive and negative helicity during each cycle; filled circles are the numerically calculated values. The orange, red, and blue curves represent the $k=1$, $k=0.75$, and $k=0.5$ cases, respectively.}
\label{fig:helicityf}
\end{figure*}

\begin{figure*}
\centering\includegraphics[scale=0.4,trim=0cm 0cm 3.5cm 0cm, clip=true]{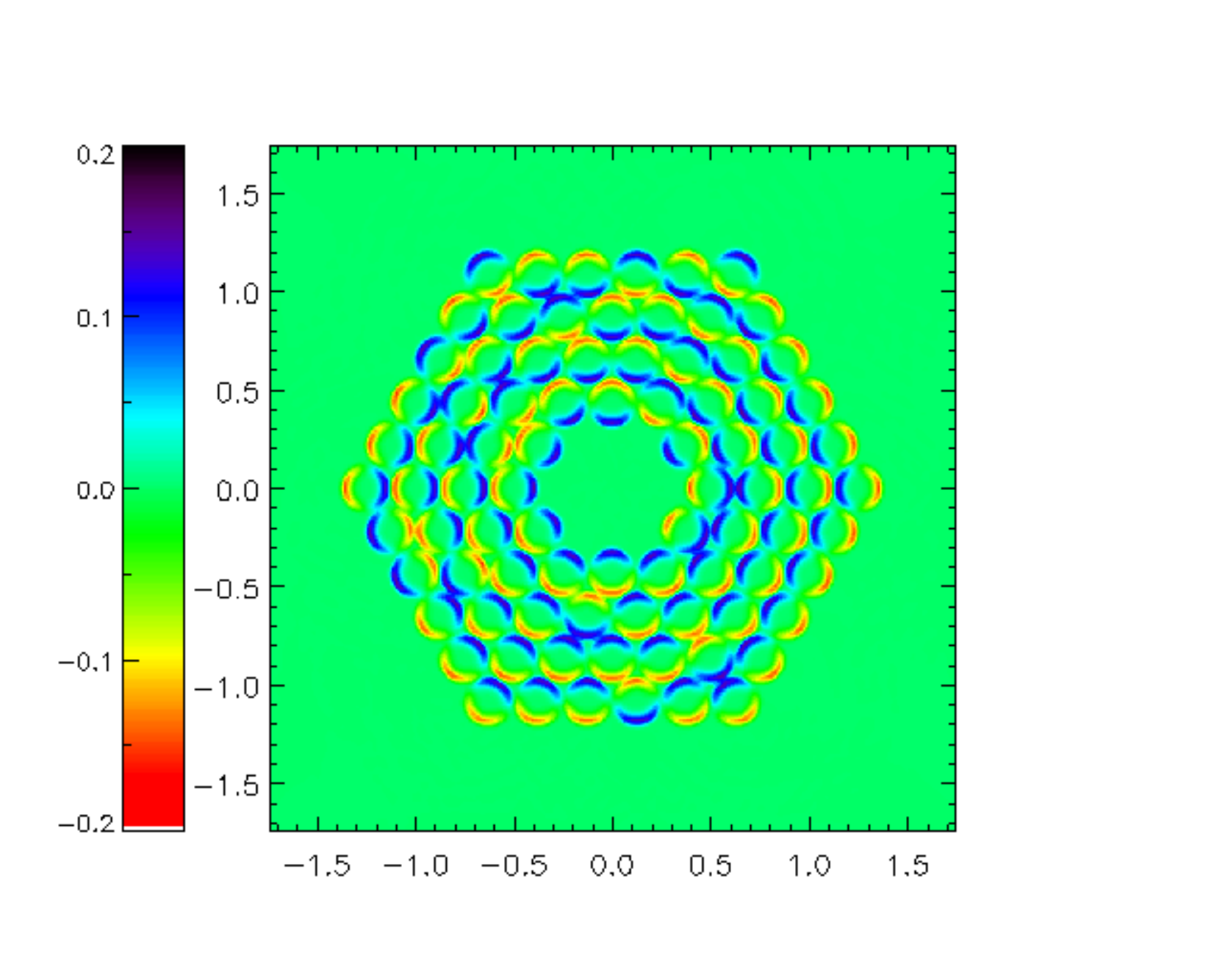}
\centering\includegraphics[scale=0.4,trim=0cm 0cm 3.5cm 0cm, clip=true]{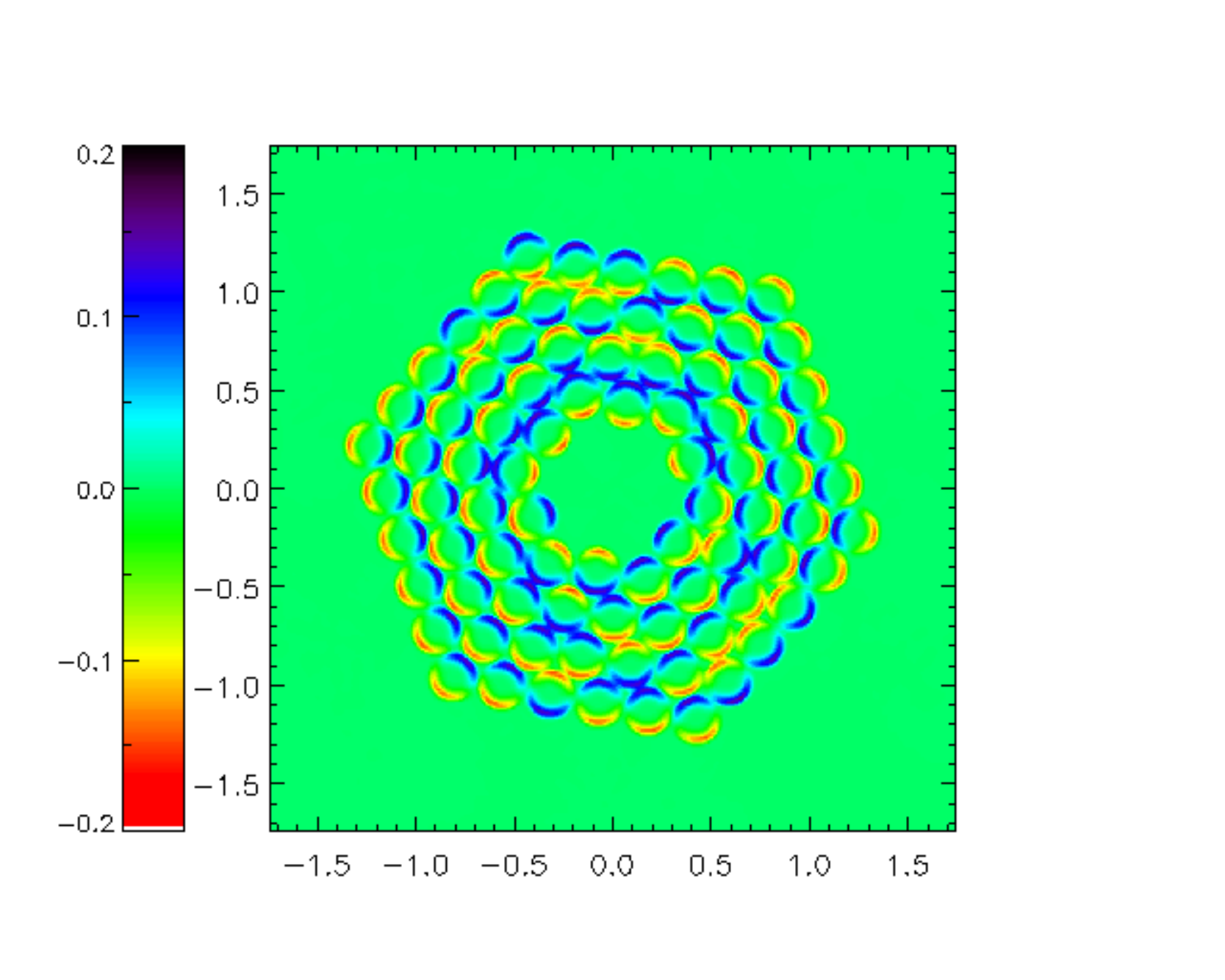}
\caption{$V_\phi$ on the bottom plate for the random-pattern case with $k=0.75$ during the first (left) and second (right) cycles.}
\label{fig:vphi_r}
\end{figure*}

\begin{figure*}[!p]
\centering\includegraphics[scale=0.75,trim=0.0cm 6.5cm 0.0cm 0.0cm]{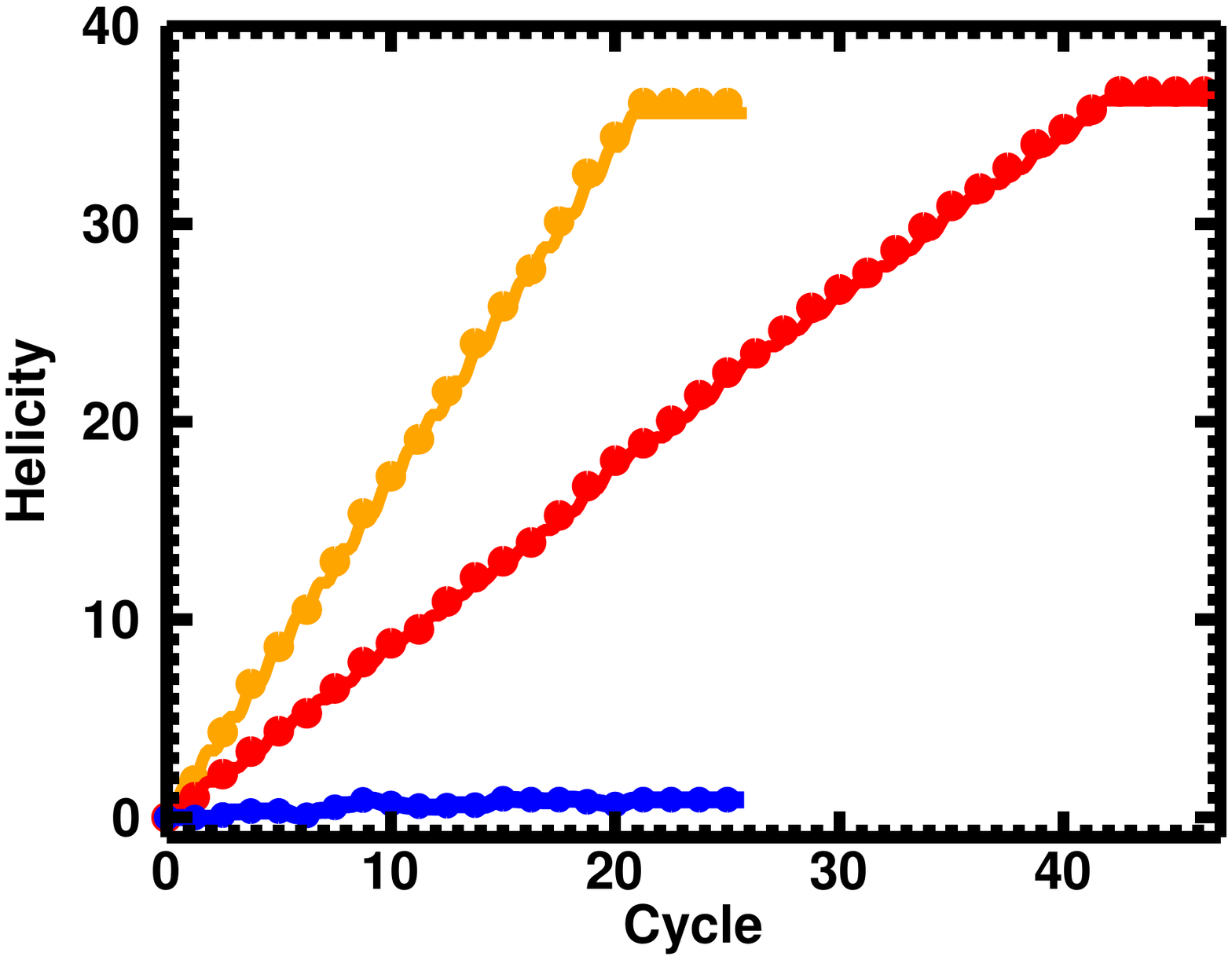}
\caption{Helicity for random-pattern simulations. Solid curves are the analytically expected helicity based on the number of convective cells injecting positive and negative helicity during each cycle; filled circles are the numerically calculated values. The orange, red, and blue curves represent the $k=1$, $k=0.75$, and $k=0.5$ cases, respectively.}
\label{fig:helicityr}
\end{figure*}

\begin{figure*}[!p]
\centering\includegraphics[scale=0.5,trim=0cm 0cm 3.5cm 0cm, clip=true]{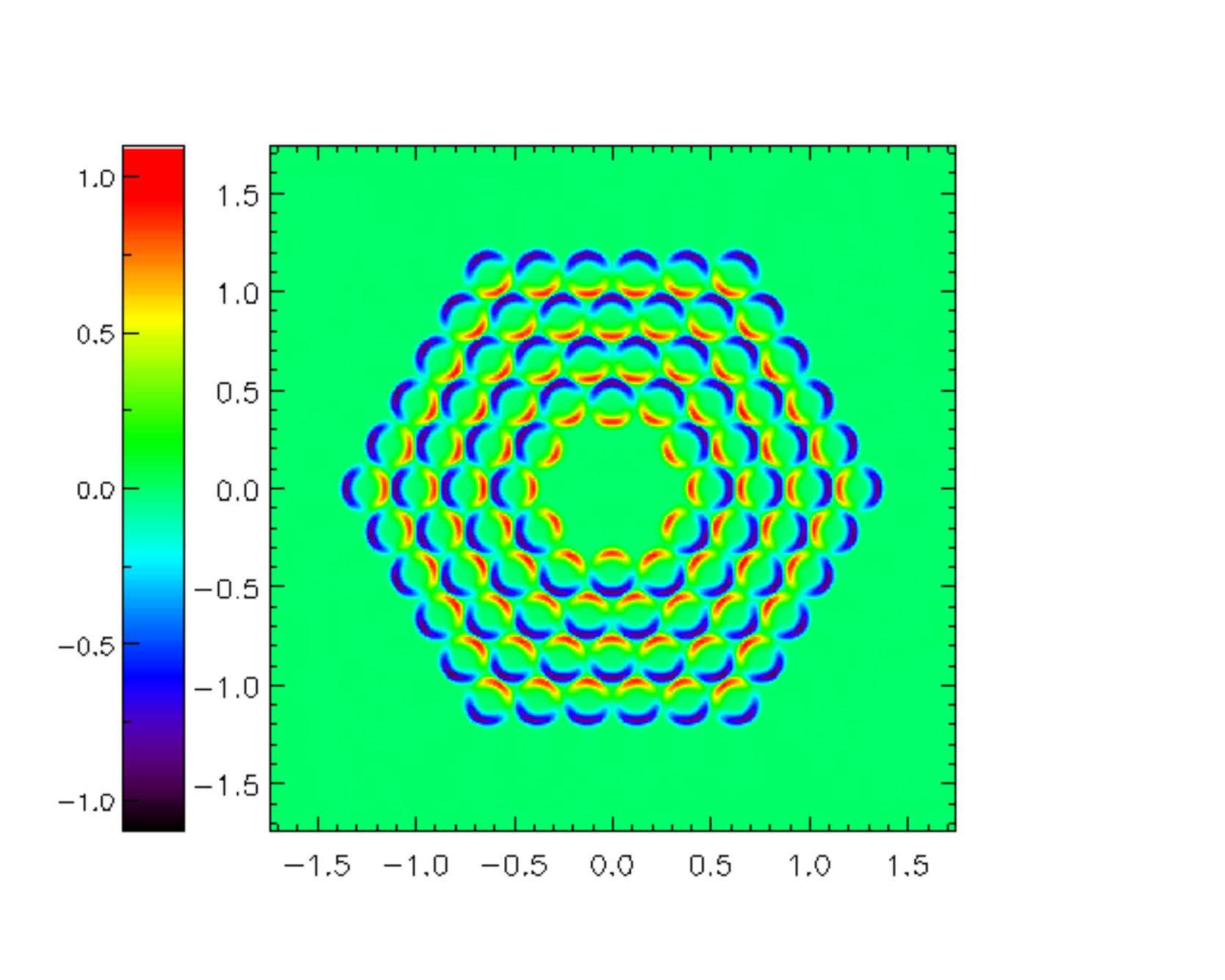}
\centering\includegraphics[scale=0.5,trim=0cm 0cm 3.5cm 0cm, clip=true]{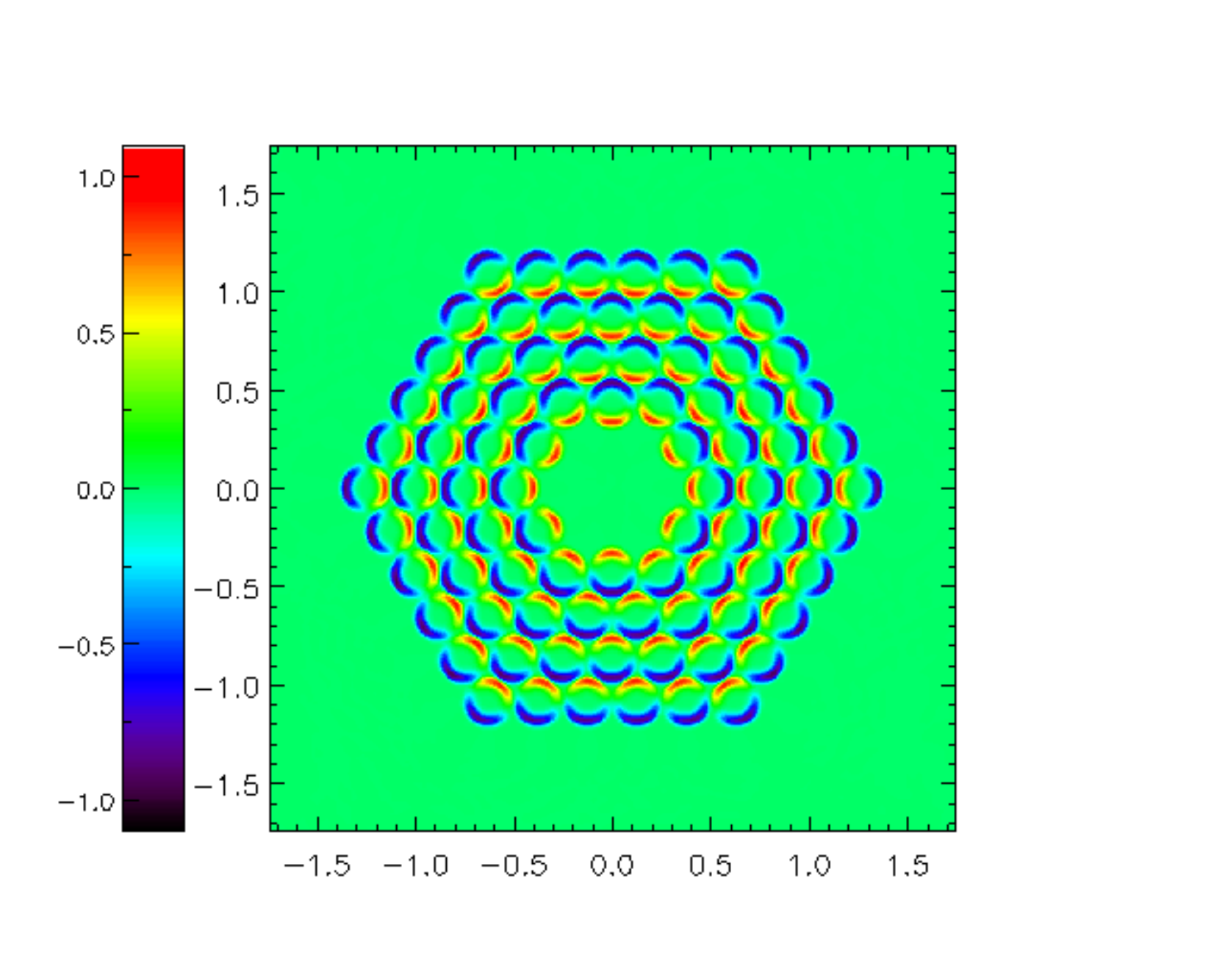}
\centering\includegraphics[scale=0.5,trim=0cm 0cm 3.5cm 0cm, clip=true]{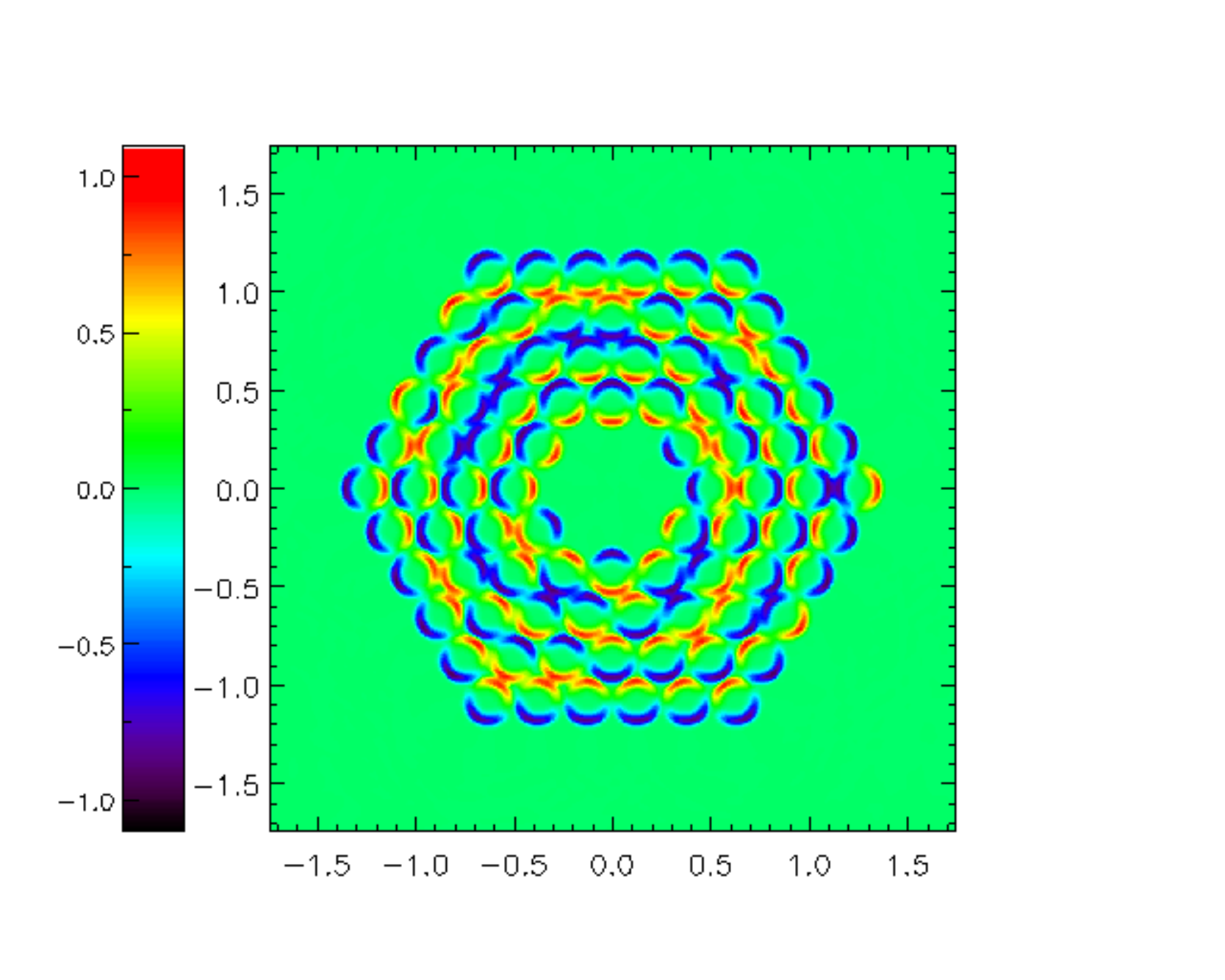}
\centering\includegraphics[scale=0.5,trim=0cm 0cm 3.5cm 0cm, clip=true]{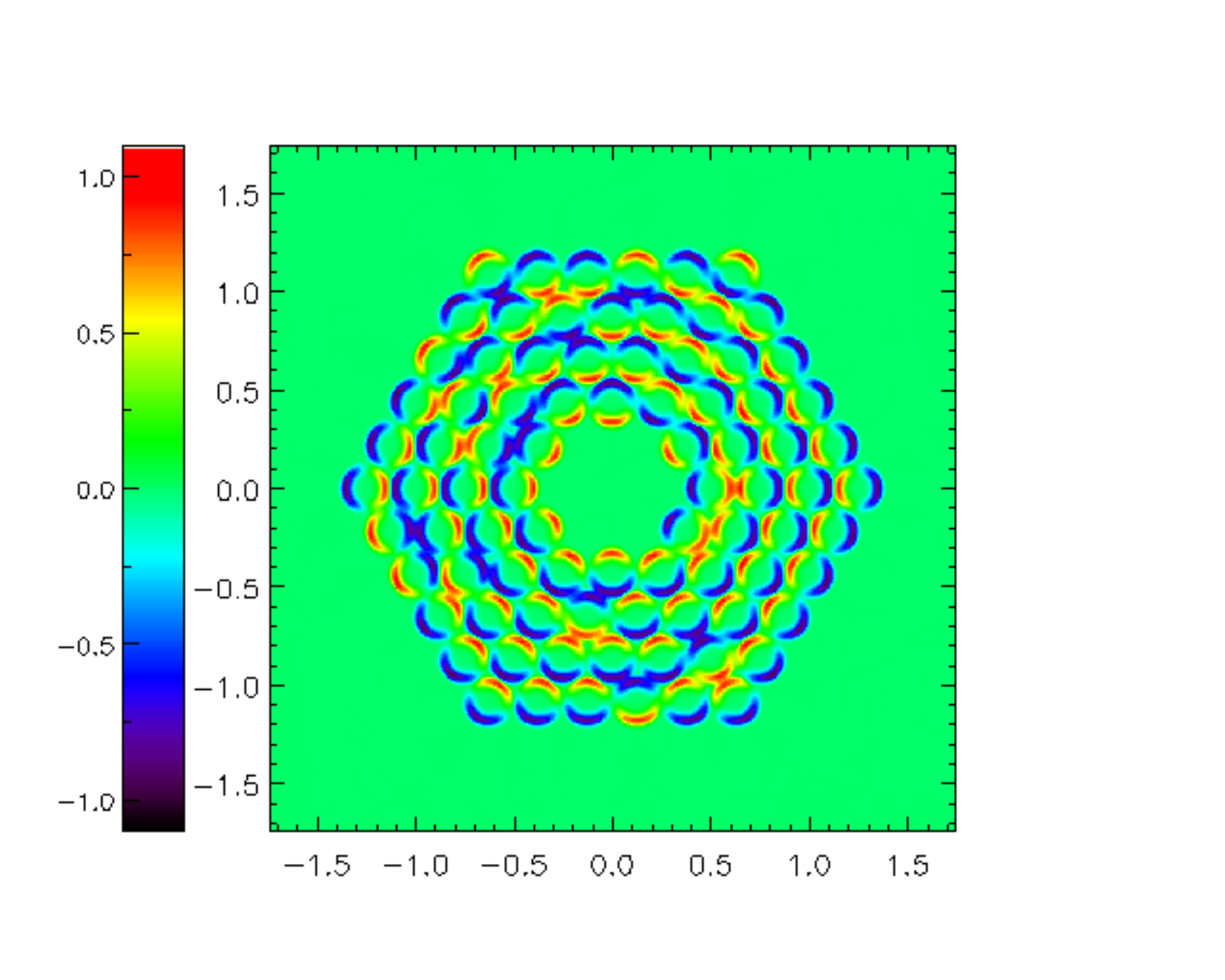}
\centering\includegraphics[scale=0.5,trim=0cm 0cm 3.5cm 0cm, clip=true]{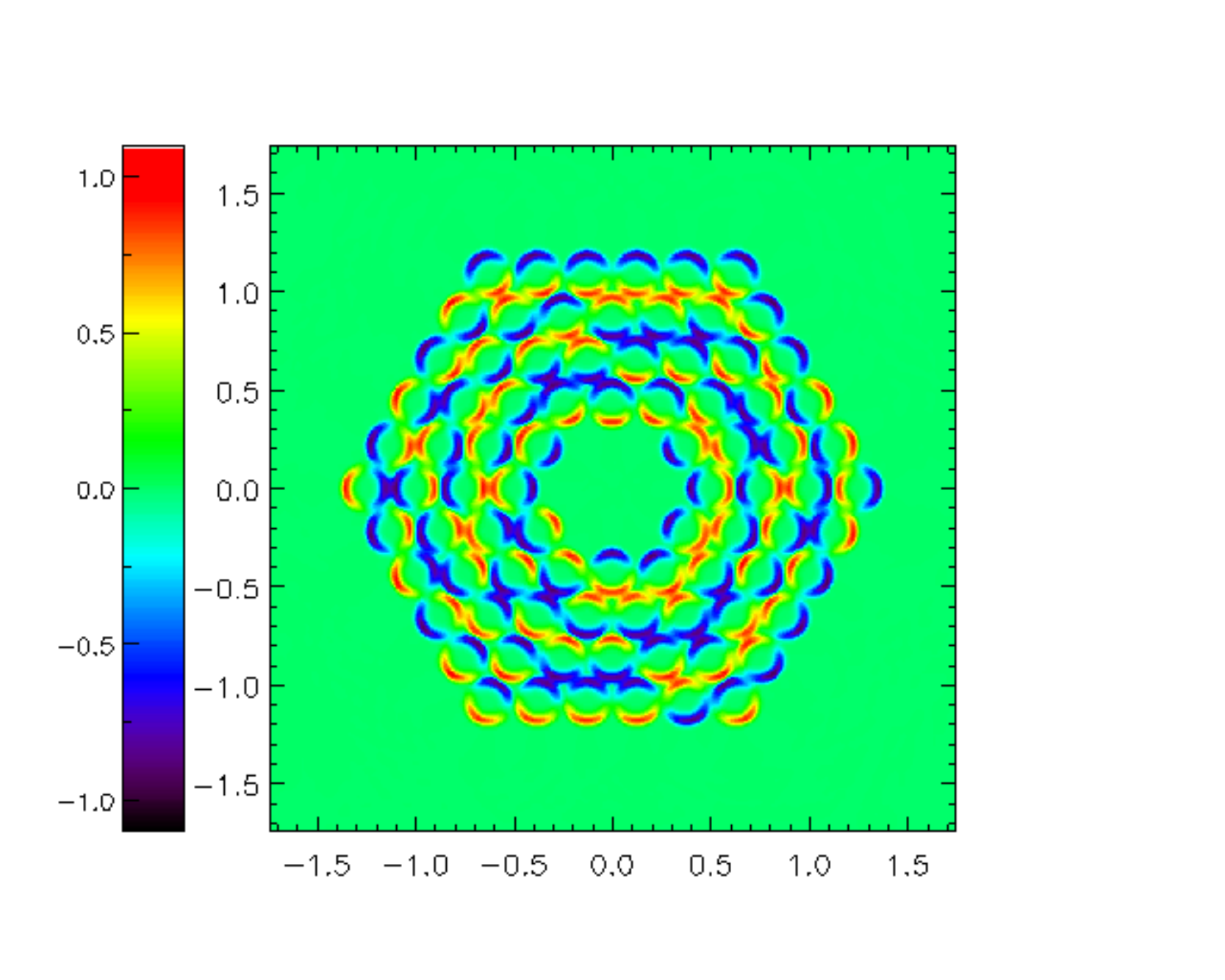}
\centering\includegraphics[scale=0.5,trim=0cm 0cm 3.5cm 0cm, clip=true]{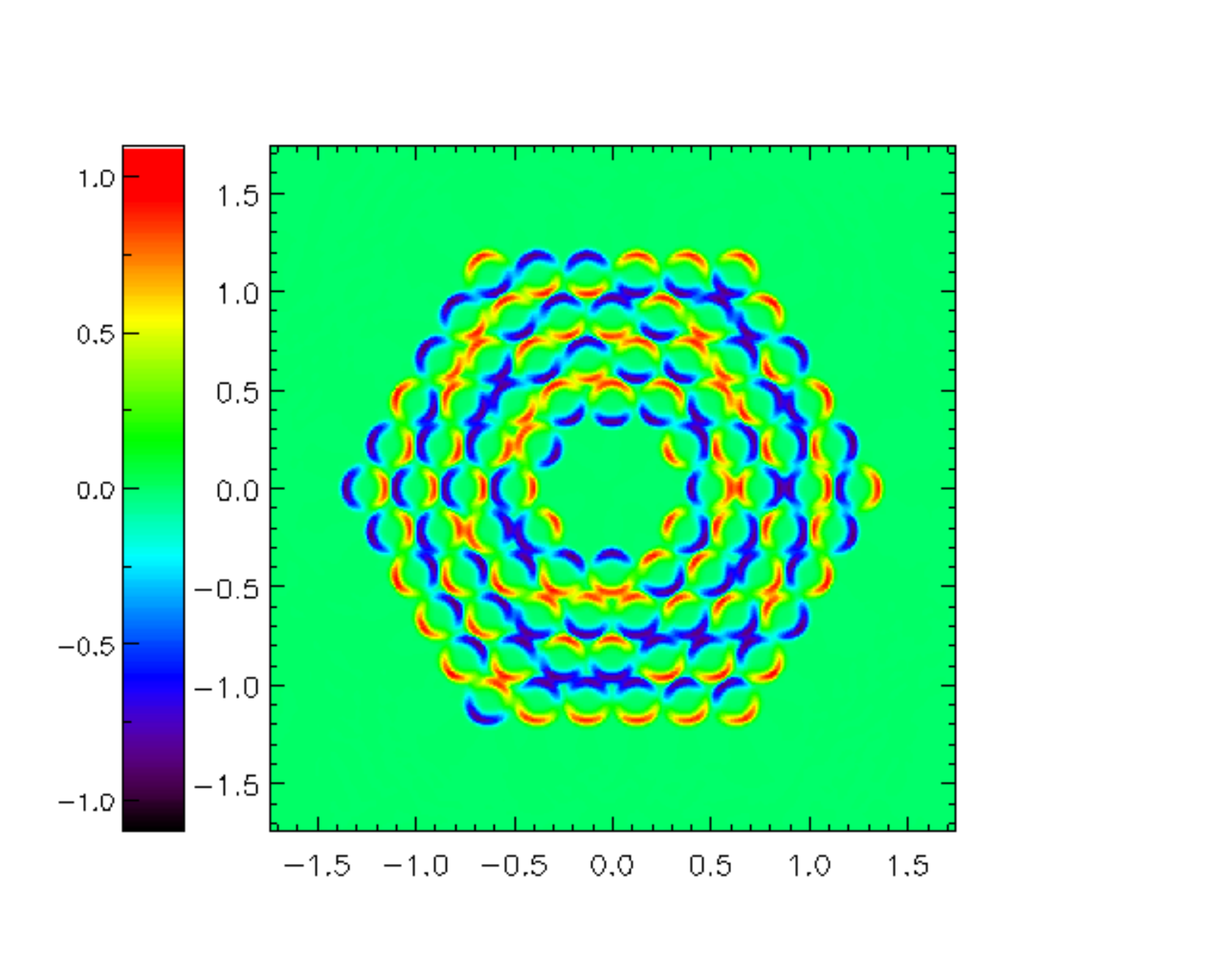}
\caption{Twist field $B_\phi$ (color shading) in the mid-plane ($x=0.5$) halfway through the first cycle of the fixed-pattern (left) and random-pattern (right) simulations. From top to bottom are the $k=1$, $k=0.75$, and $k=0.5$ cases, respectively. Red/yellow (blue/teal) represents clockwise (counter-clockwise) twist.}
\label{fig:Bphi_beg}
\end{figure*}

\begin{figure*}[!p]
\centering\includegraphics[scale=0.5,trim=0cm 0cm 3.5cm 0cm, clip=true]{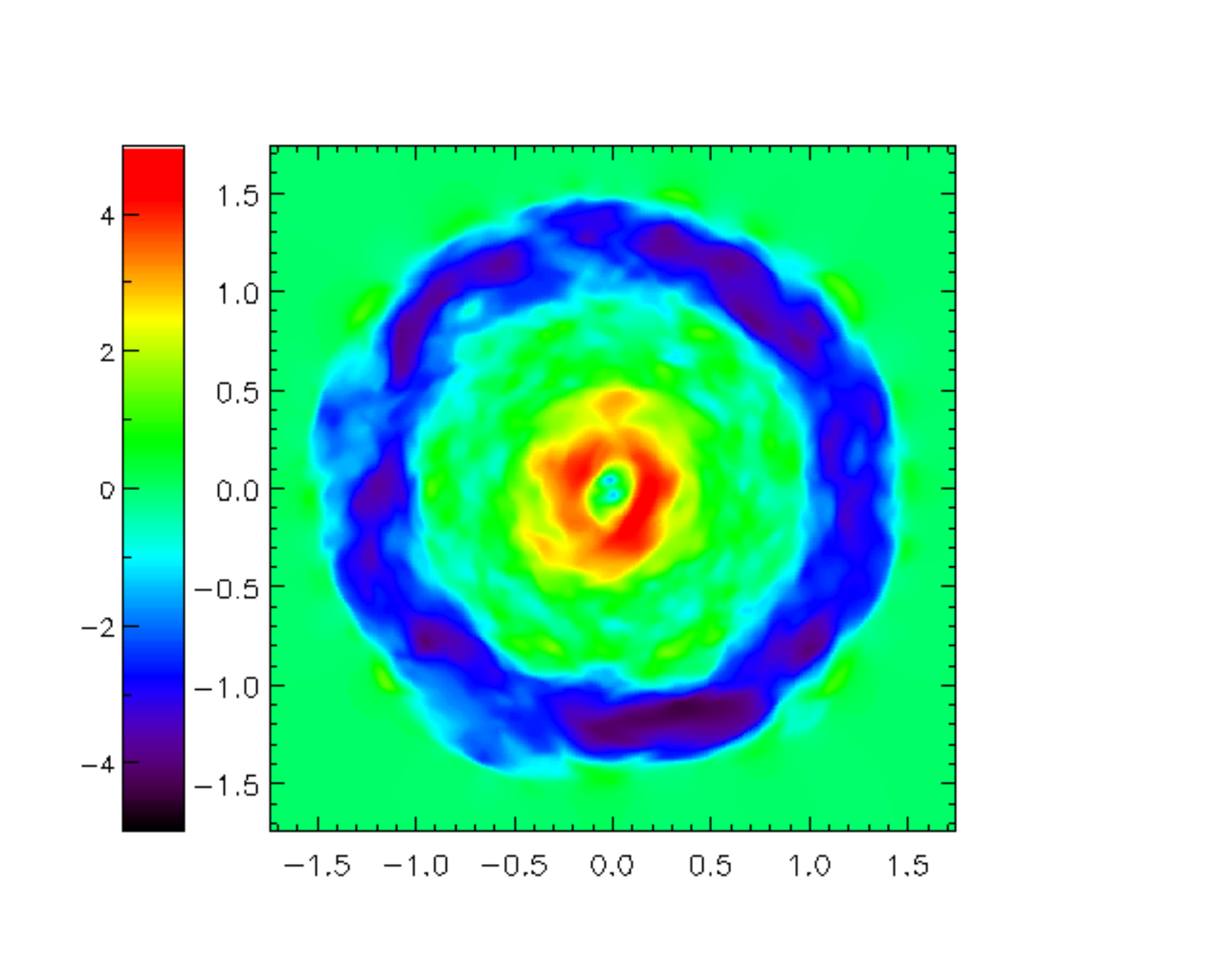}
\centering\includegraphics[scale=0.5,trim=0cm 0cm 3.5cm 0cm, clip=true]{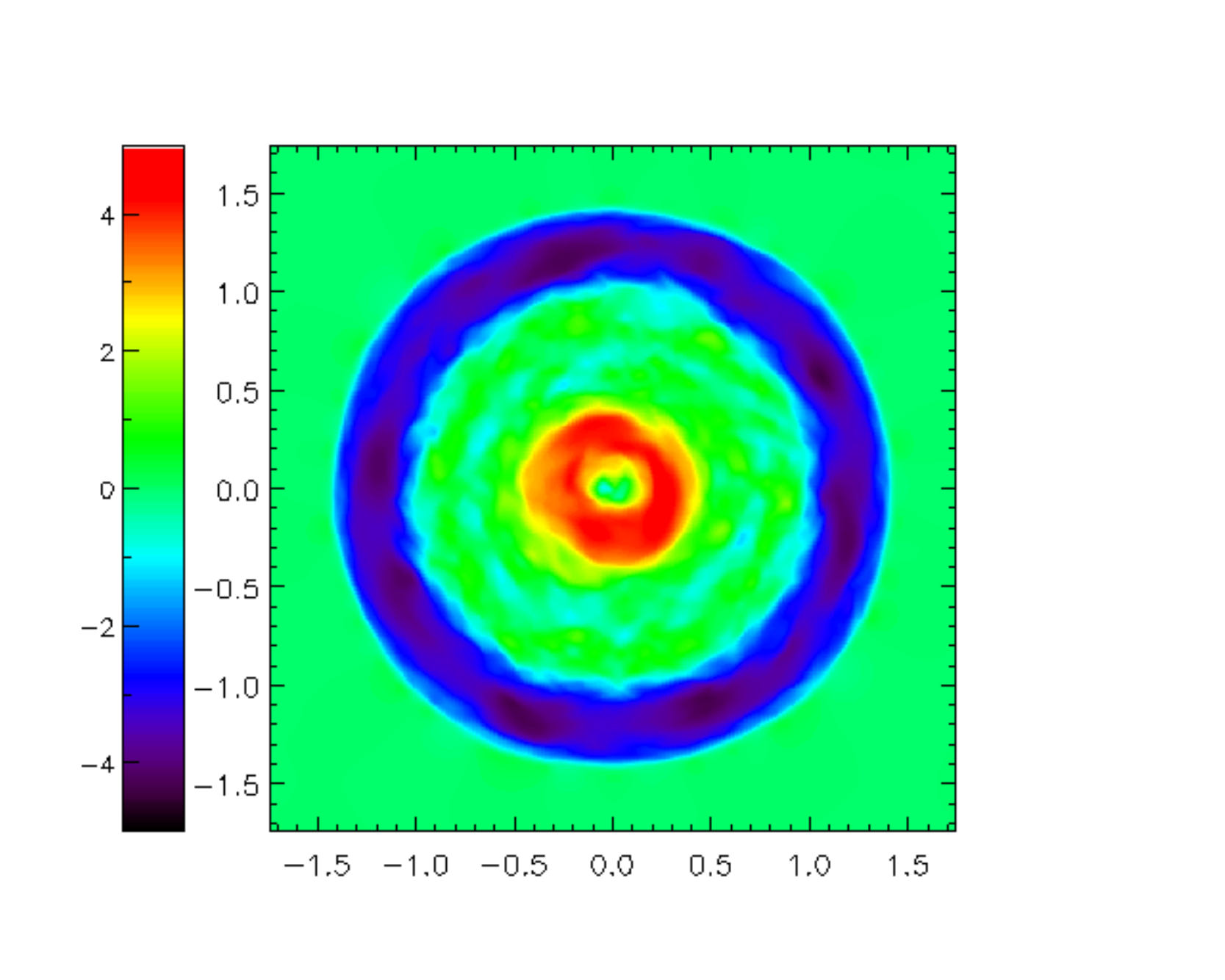}
\centering\includegraphics[scale=0.5,trim=0cm 0cm 3.5cm 0cm, clip=true]{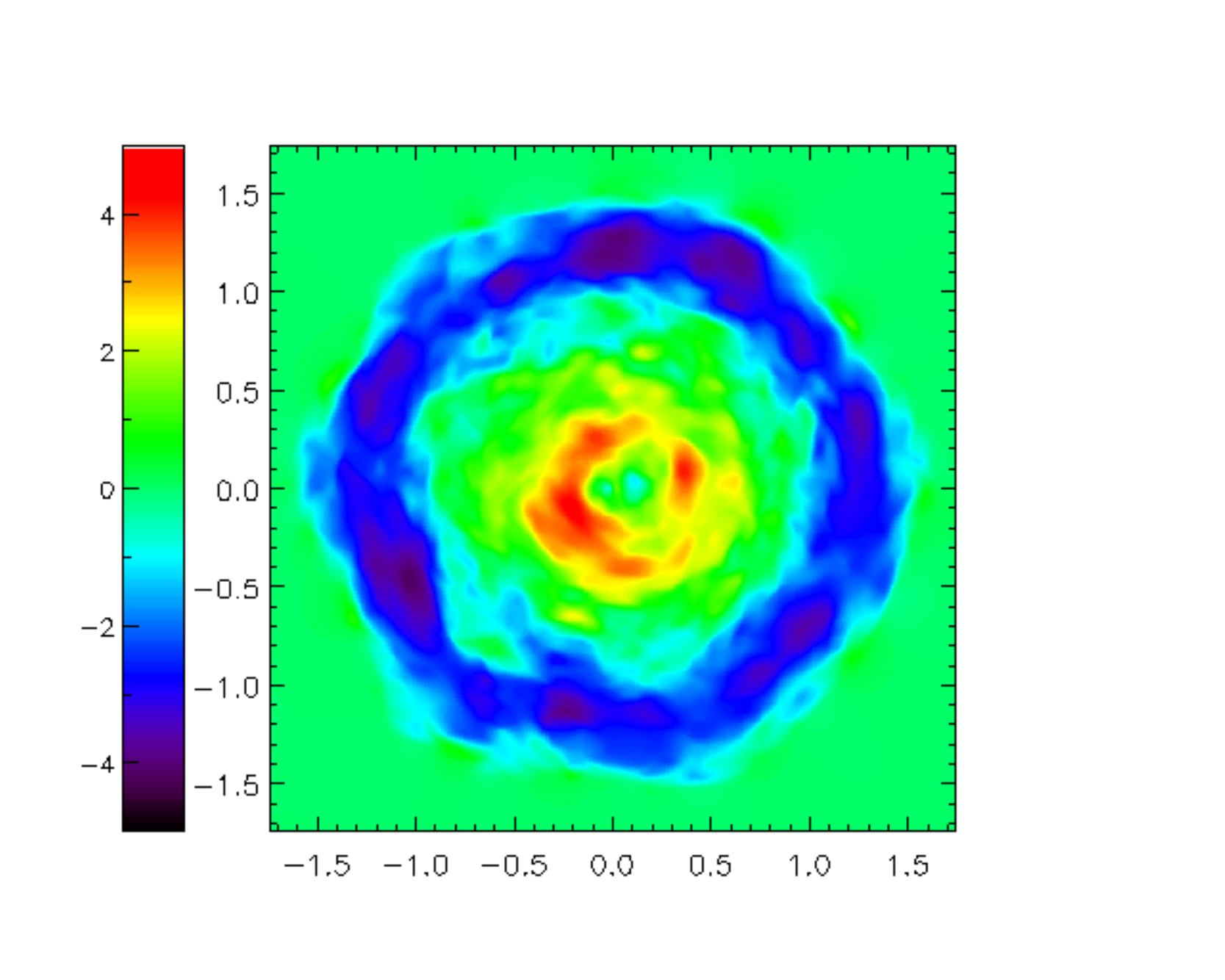}
\centering\includegraphics[scale=0.5,trim=0cm 0cm 3.5cm 0cm, clip=true]{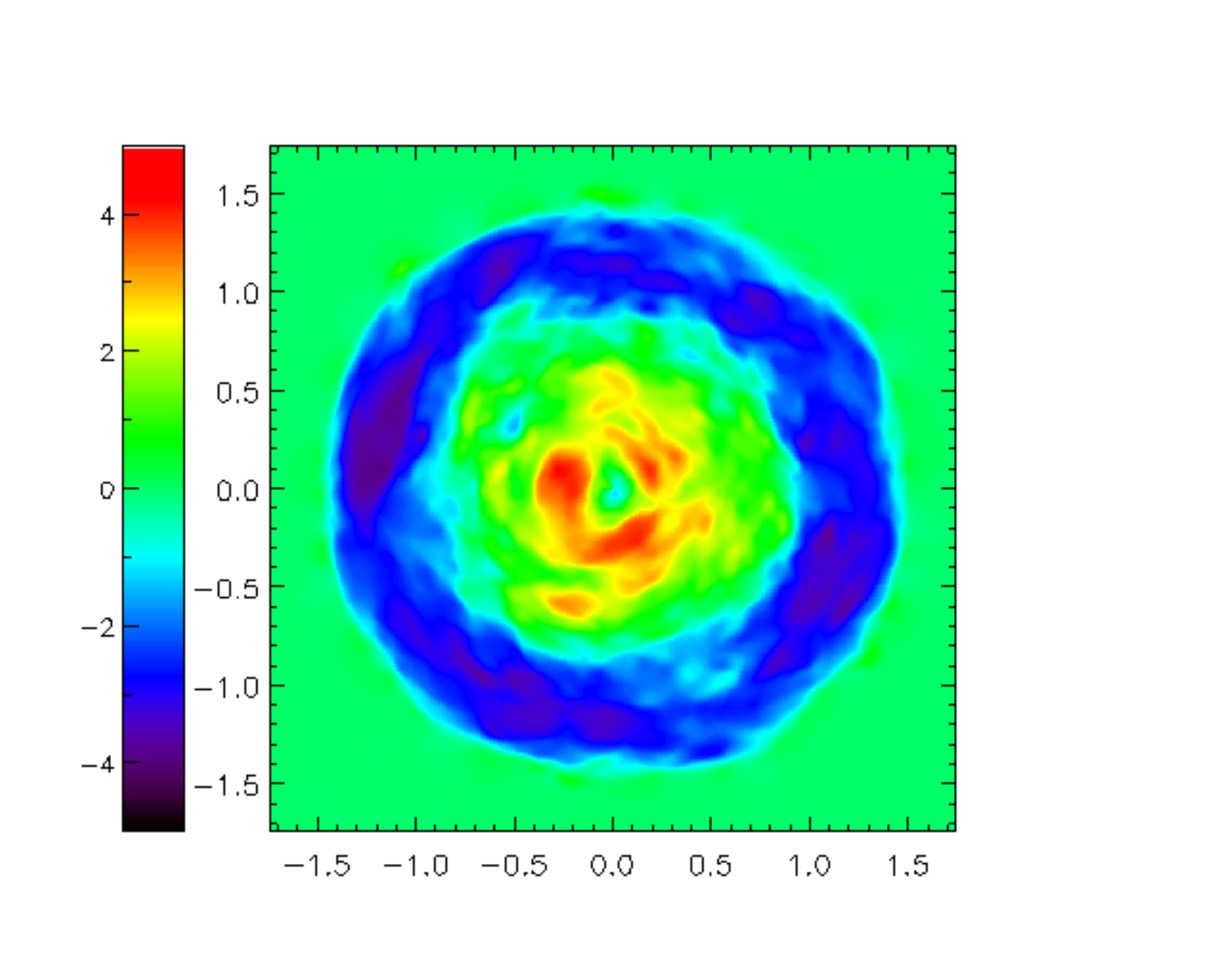}
\centering\includegraphics[scale=0.5,trim=0cm 0cm 3.5cm 0cm, clip=true]{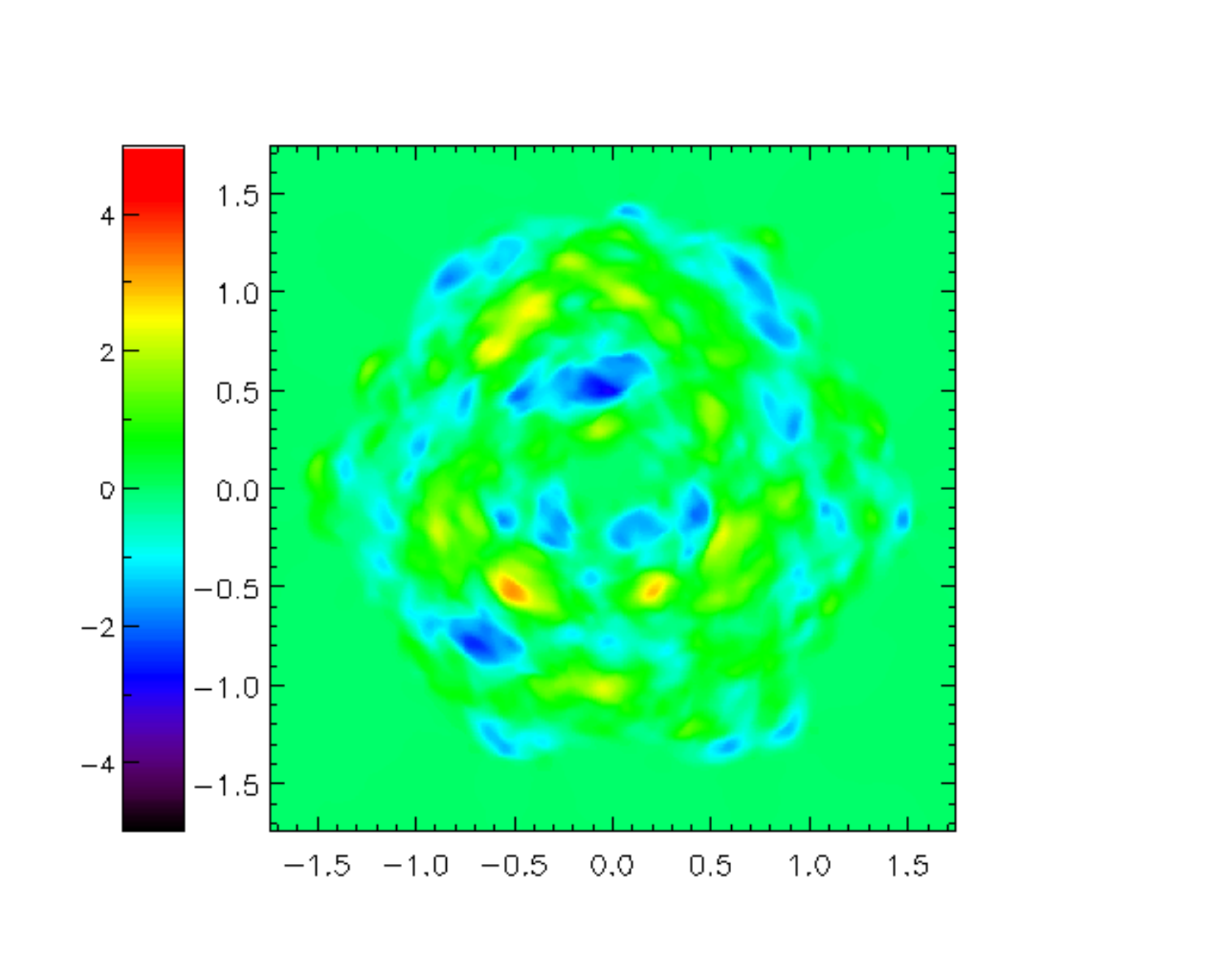}
\centering\includegraphics[scale=0.5,trim=0cm 0cm 3.5cm 0cm, clip=true]{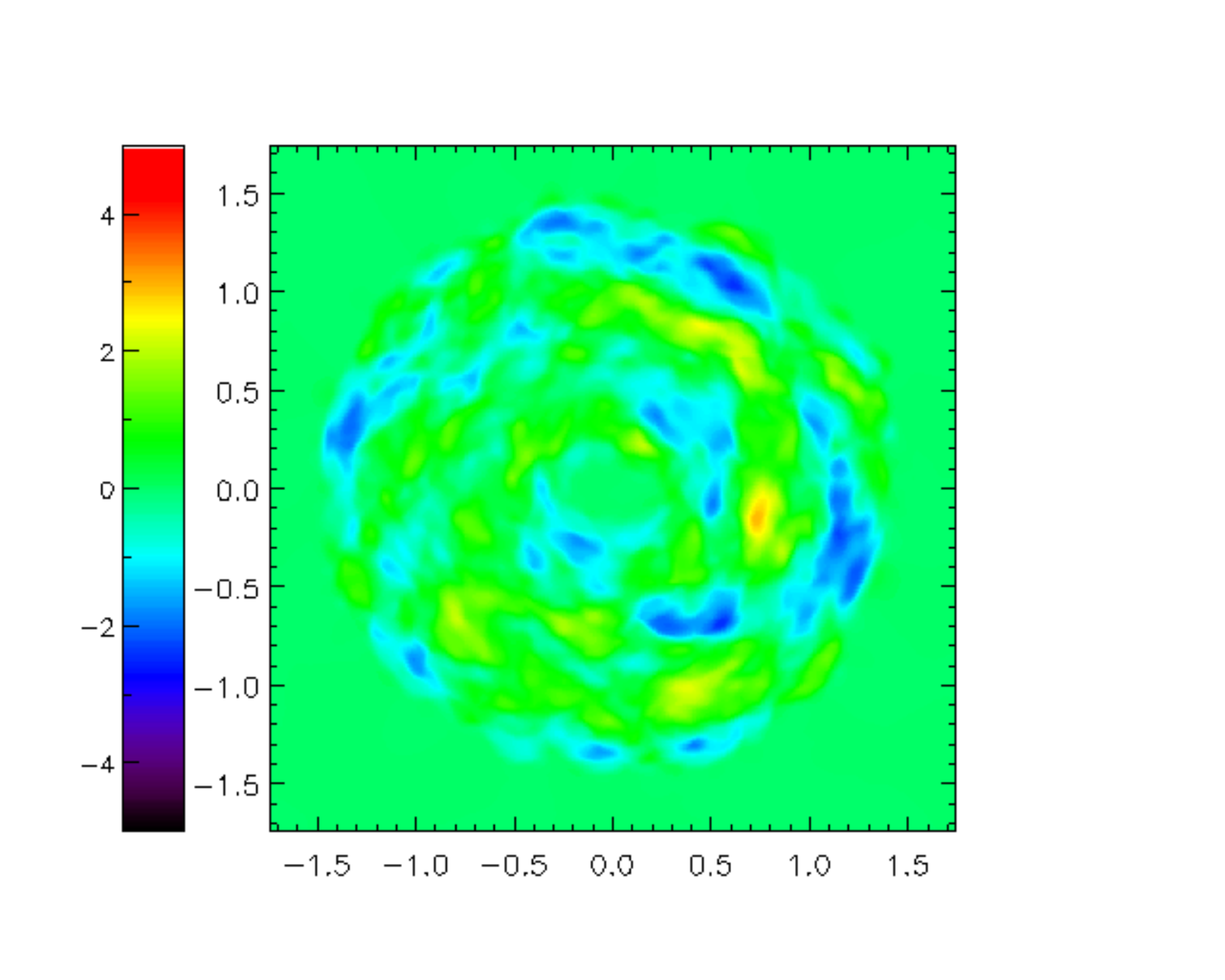}
\caption{Twist field $B_\phi$ (color shading) in the mid-plane ($x=0.5$) at the end of the fixed-pattern (left) and random-pattern (right) simulations. From top to bottom are the $k=1$, $k=0.75$, and $k=0.5$ cases, respectively. Red/yellow (blue/teal) represents clockwise (counter-clockwise) twist.}
\label{fig:Bphi_end}
\end{figure*}

\begin{figure*}
\centering\includegraphics[scale=0.75, trim=0.0cm 6.5cm 0.0cm 0.0cm]{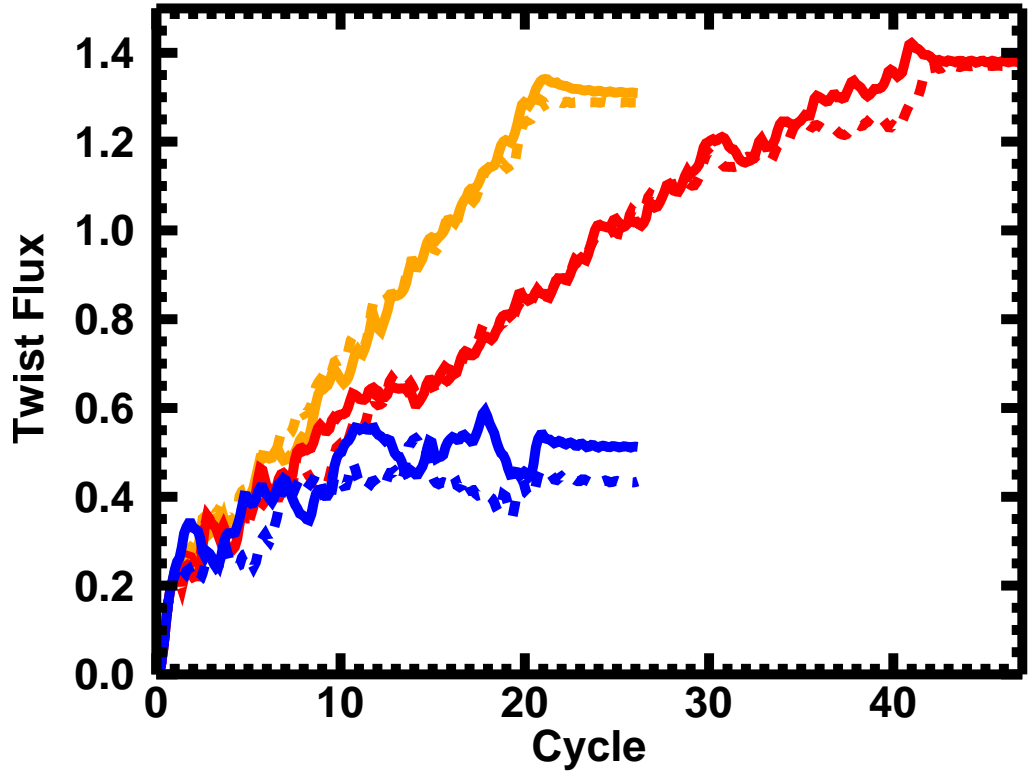}
\caption{Positive twist flux $\Phi_{tw}^+$ versus cycle through the vertical half-plane $z=0$ for the fixed-pattern (solid curves) and random-pattern (dashed curves) cases with $k=1$ (orange), $k=0.75$ (red), and $k=0.5$ (blue).}
\label{fig:tw_v_time}
\end{figure*}

\begin{figure*}[!h]
\centering\includegraphics[scale=0.35, trim=0.0cm 0.0cm 0.0cm 2.2cm,clip=true]{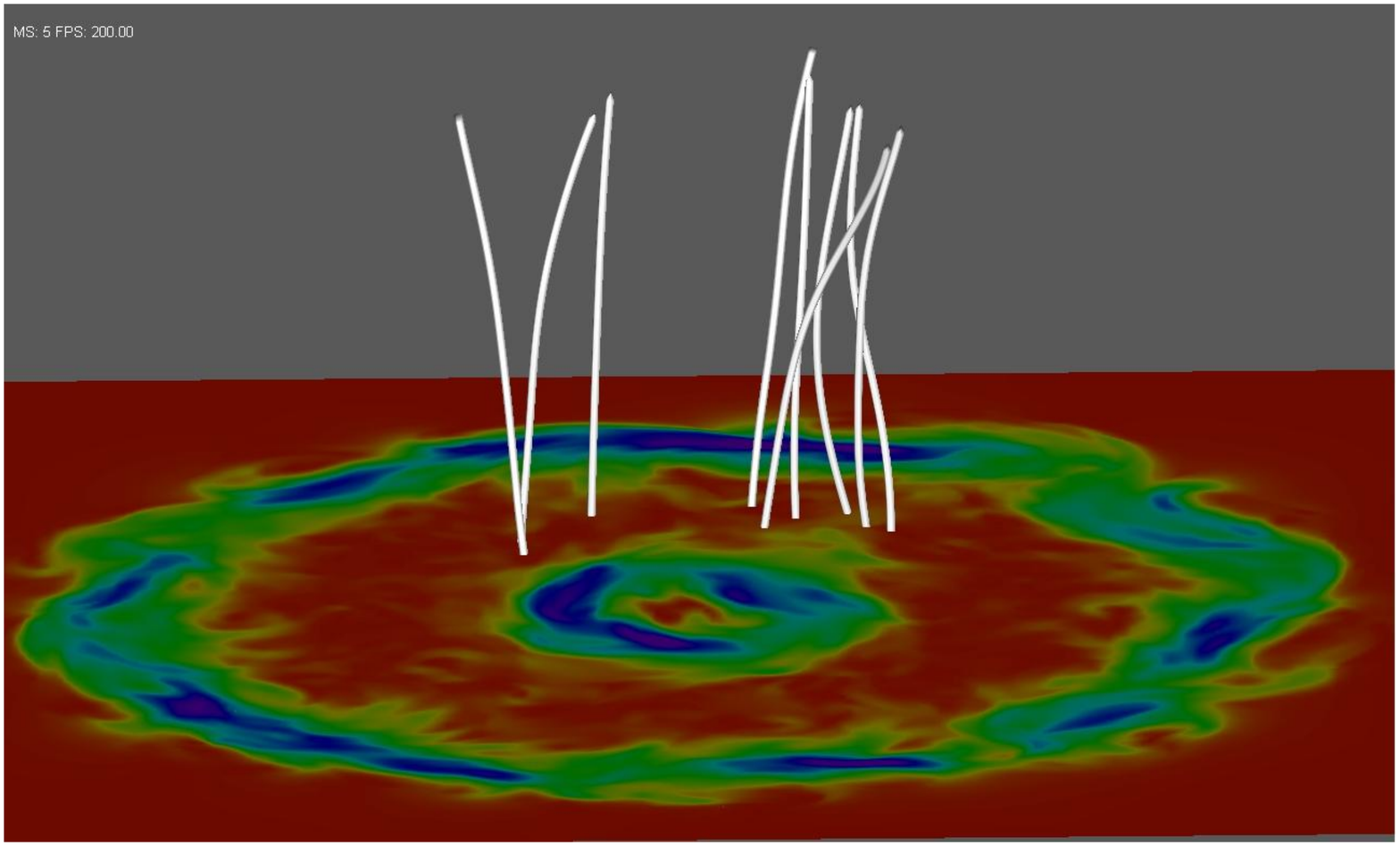}
\centering\includegraphics[scale=0.35, trim=0.0cm 0.0cm 0.0cm 2.2cm,clip=true]{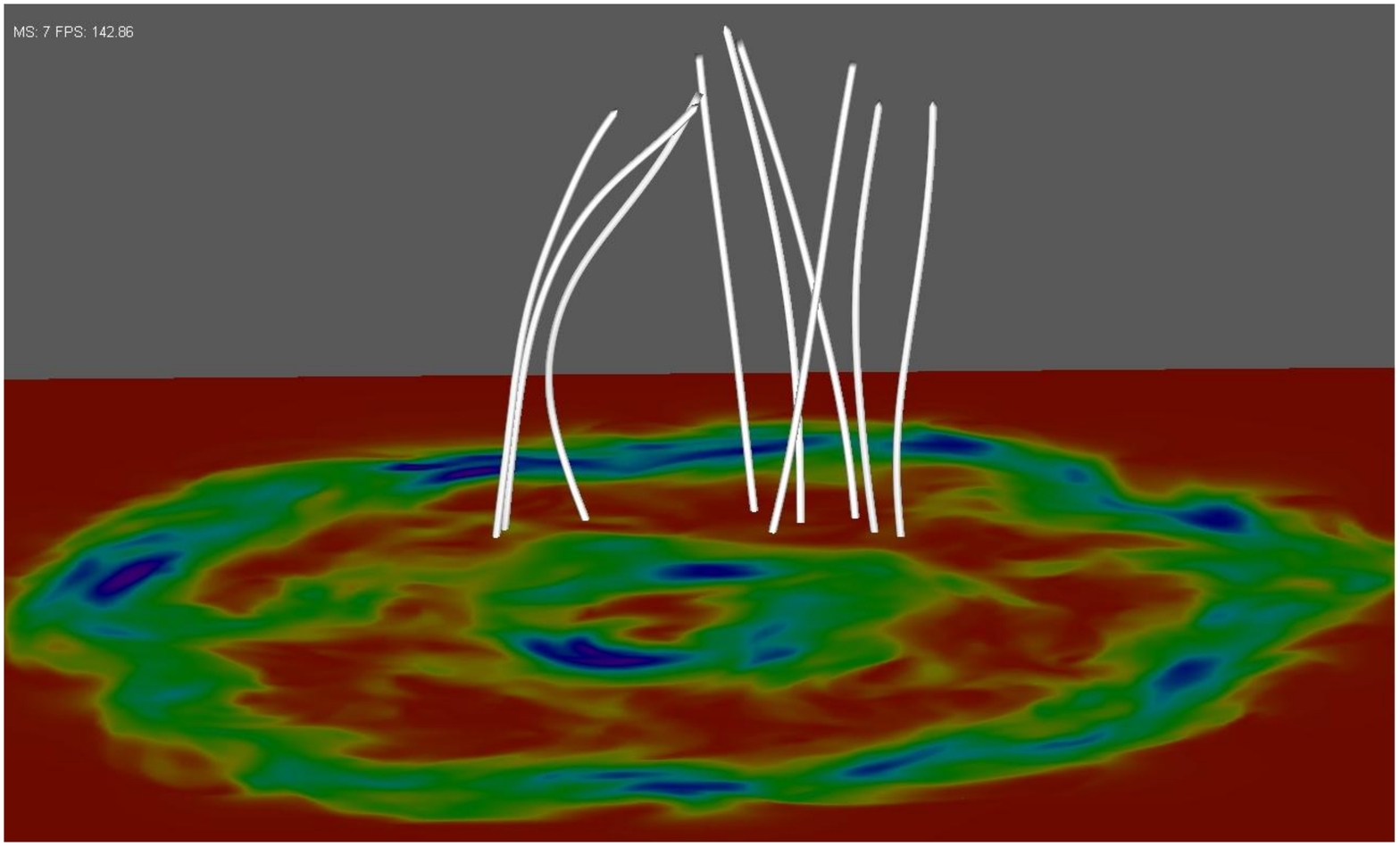}
\centering\includegraphics[scale=0.35, trim=0.0cm 0.0cm 0.0cm 2.2cm,clip=true]{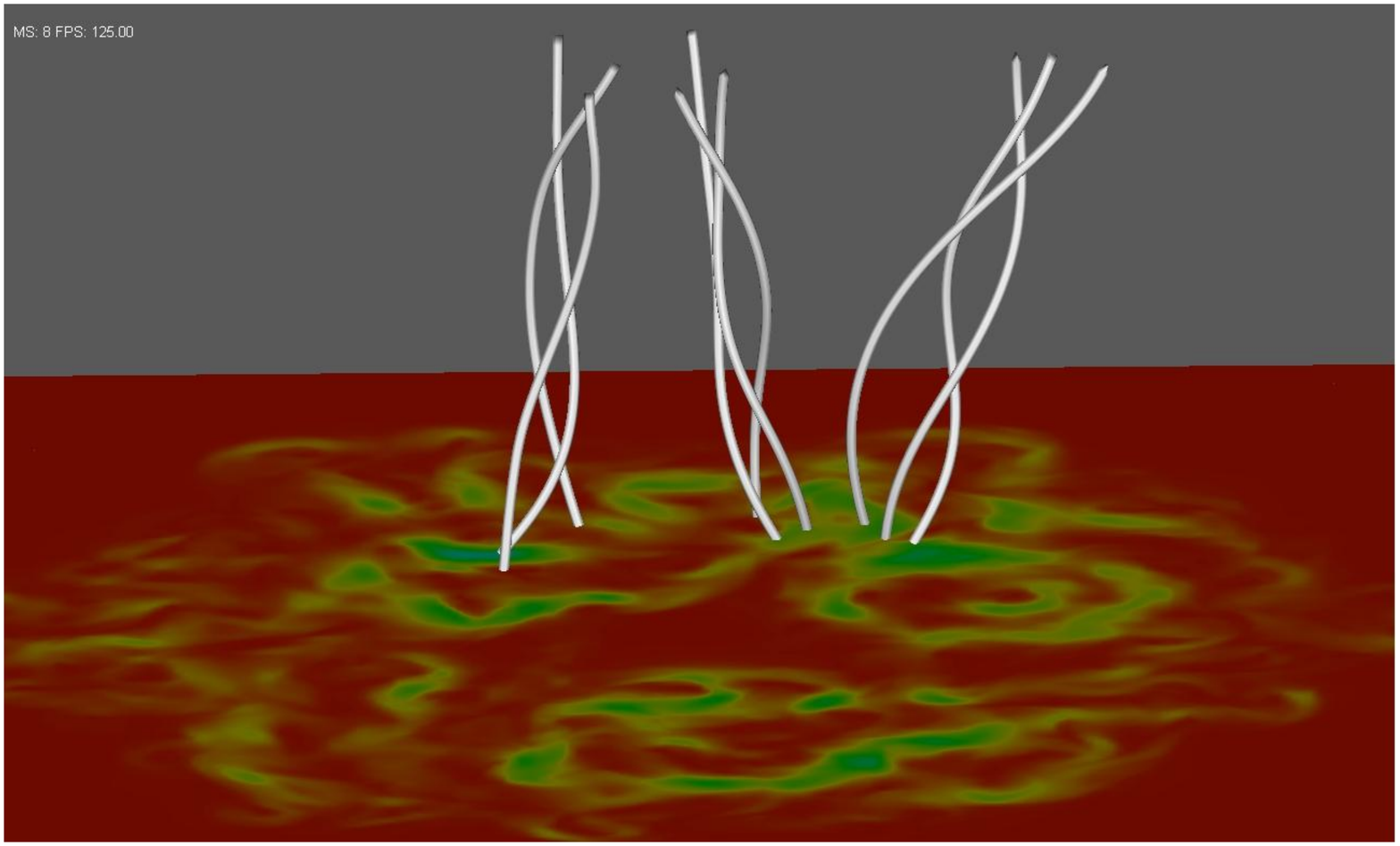}
\centering\includegraphics[scale=0.35, trim=0.0cm 7.0cm 0.0cm  9.2cm,clip=true]{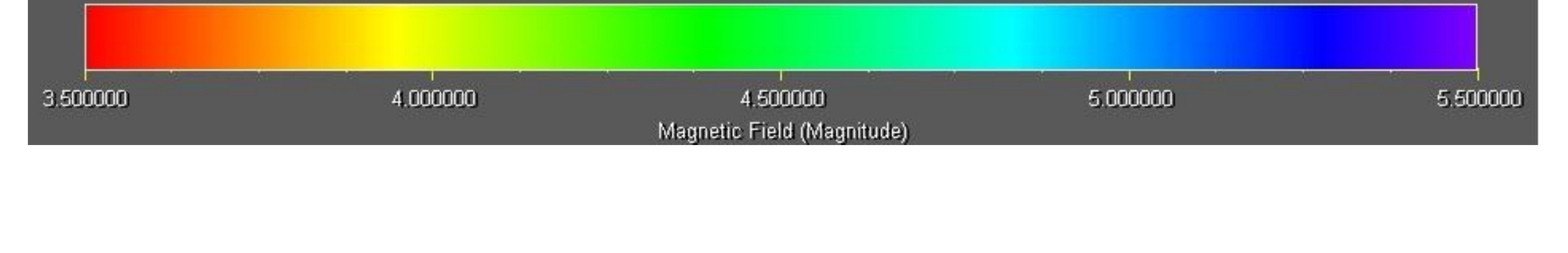}
\caption{ A sample of `coronal loop' field lines (white) plotted at the end of the fixed-pattern $k=1$ (top) $k=0.75$ (middle) and $k=0.5$ (bottom) simulations. The field lines are plotted inside the interior of the hexagonal region on the bottom plate, which shows magnetic field magnitude (color shading). The initial (potential) field is a uniform straight field.}
\label{fig:structure_f}
\end{figure*}

\begin{figure*}[!h]
\centering\includegraphics[scale=0.35, trim=0.0cm 0.0cm 0.0cm 2.2cm,clip=true]{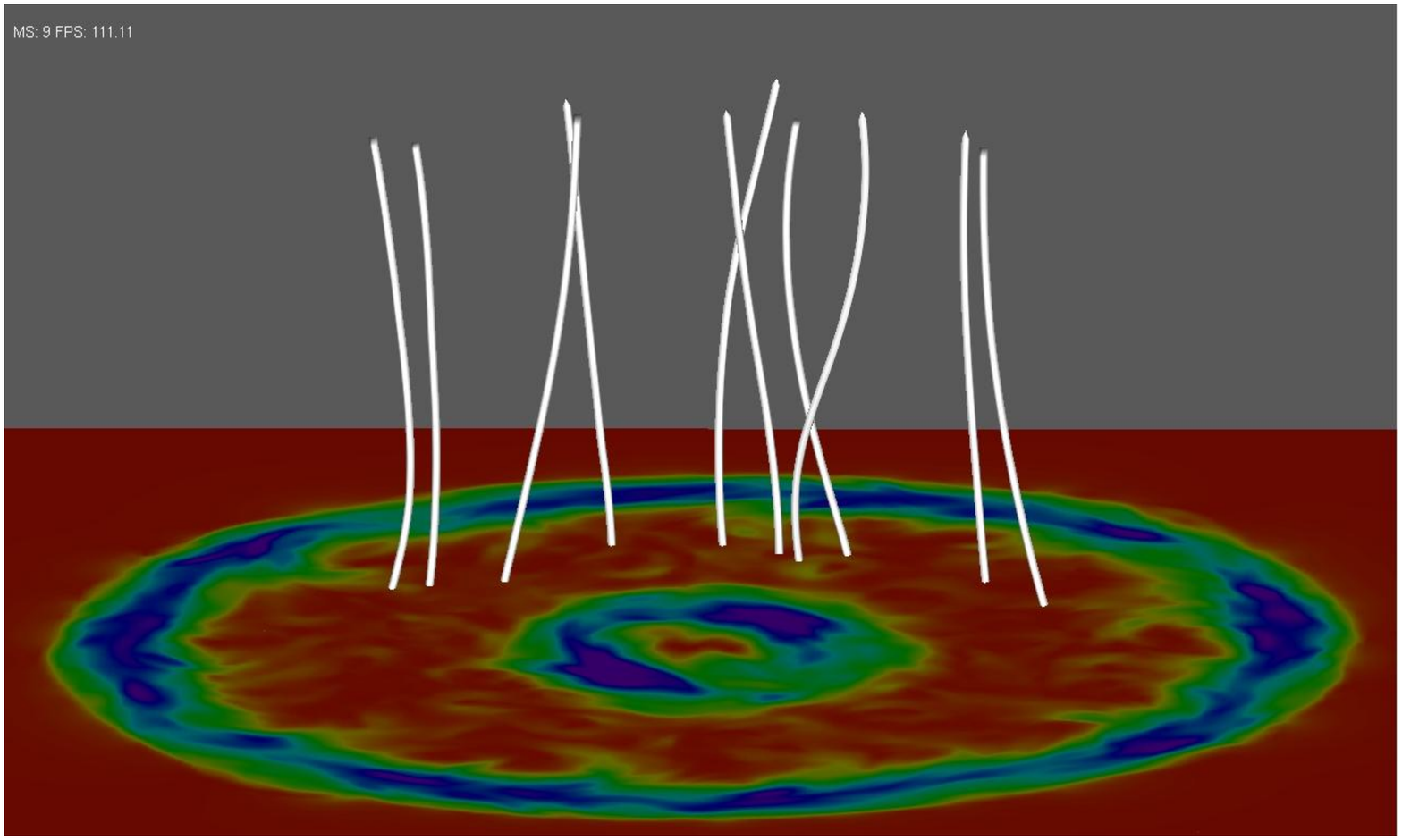}
\centering\includegraphics[scale=0.35, trim=0.0cm 0.0cm 0.0cm 2.2cm,clip=true]{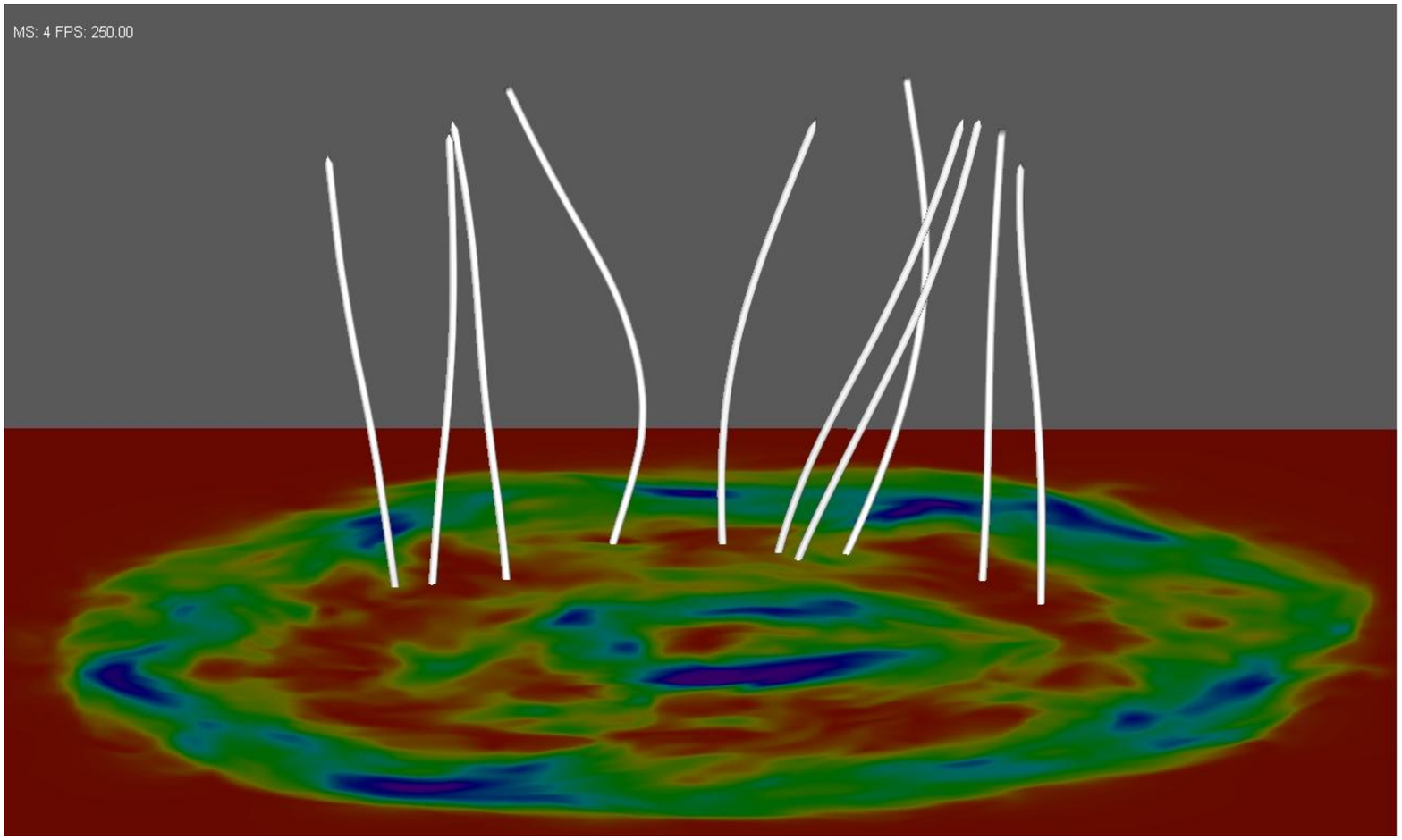}
\centering\includegraphics[scale=0.35, trim=0.0cm 0.0cm 0.0cm 2.2cm,clip=true]{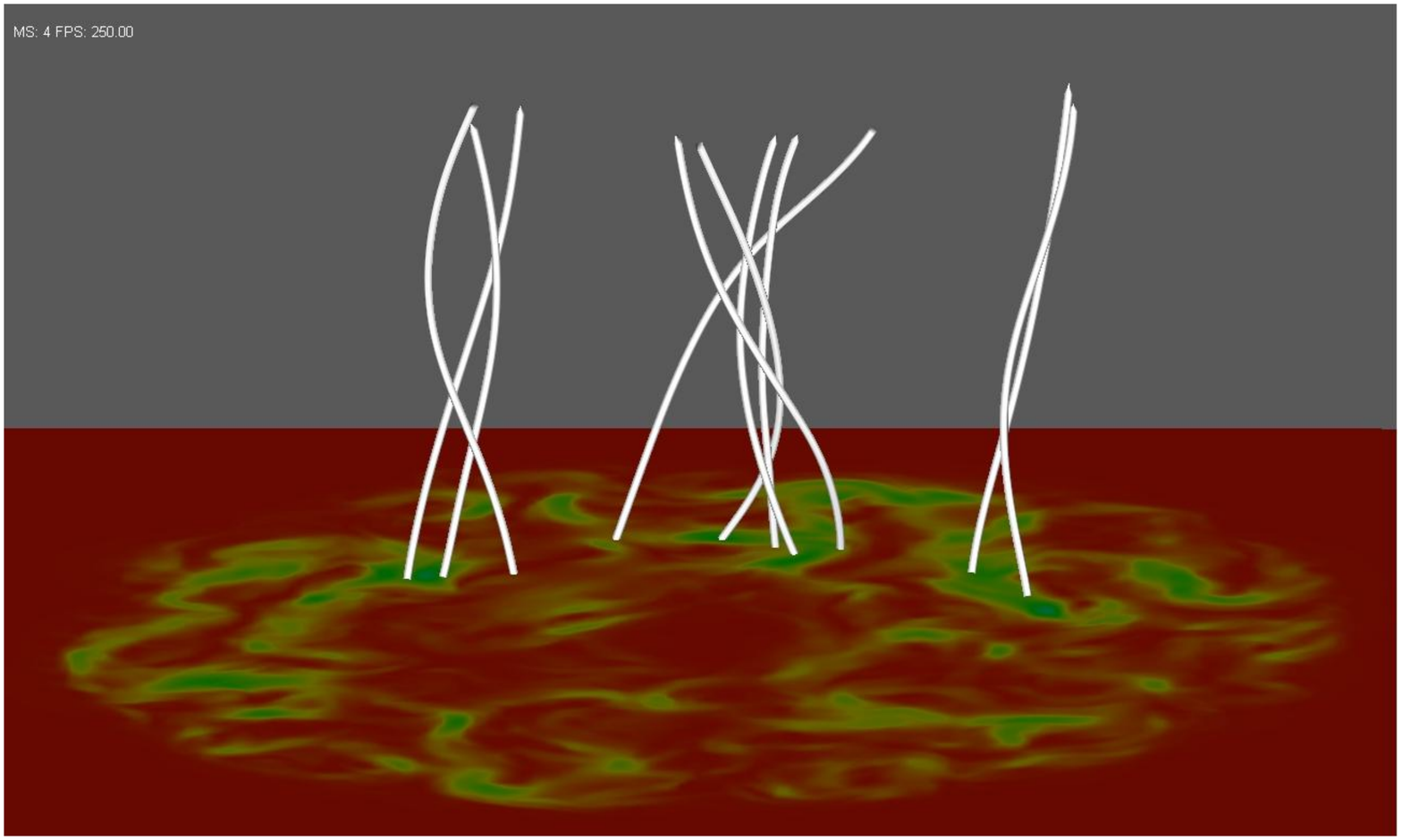}
\centering\includegraphics[scale=0.35, trim=0.0cm 7.0cm 0.0cm 9.2cm,clip=true]{colorbar.pdf}
\caption{A sample of `coronal loop' field lines (white) plotted at the end of the random-pattern $k=1$ (top) $k=0.75$ (middle) and $k=0.5$ (bottom) simulations. The field lines are plotted inside the interior of the hexagonal region on the bottom plate, which shows magnetic field magnitude (color shading). The initial (potential) field is a uniform straight field.}
\label{fig:structure_r}
\end{figure*}

\begin{figure}[!p]
\centering\includegraphics[scale=0.65, trim=0.0cm 7.0cm 0.0cm 0.0cm]{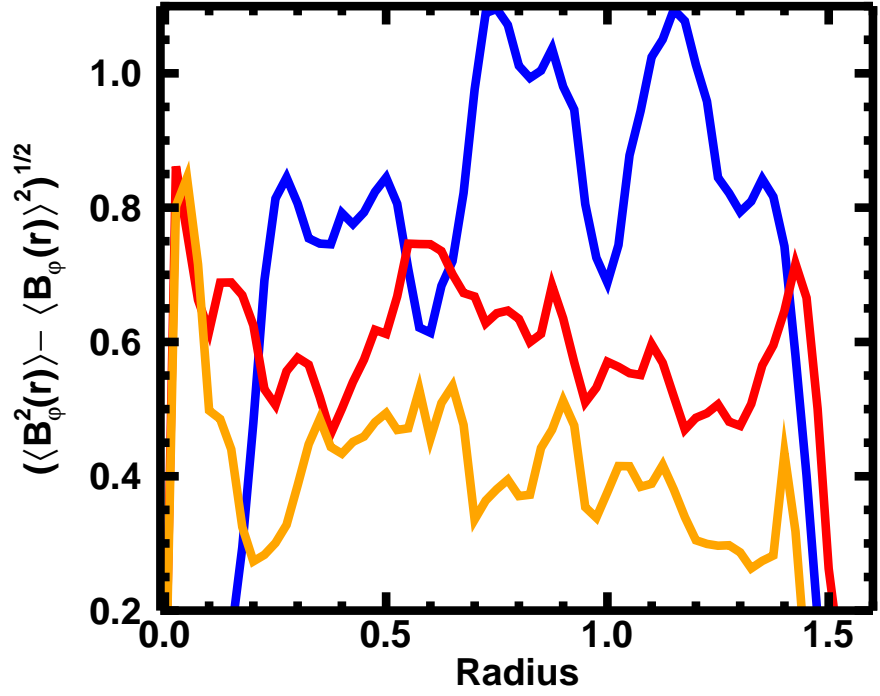}
\caption{Root mean square deviation of azimuthal magnetic field $\delta B_\phi(r)$ versus radius in the mid-plane ($x=0.5$) at the end of the random-pattern case with $k=1$ (orange), $k=0.75$ (red), and $k=0.5$ (blue).}
\label{fig:width}
\end{figure}

\begin{figure*}[!p]
\centering\includegraphics[scale=0.5,trim=0cm 0cm 3.5cm 0cm, clip=true]{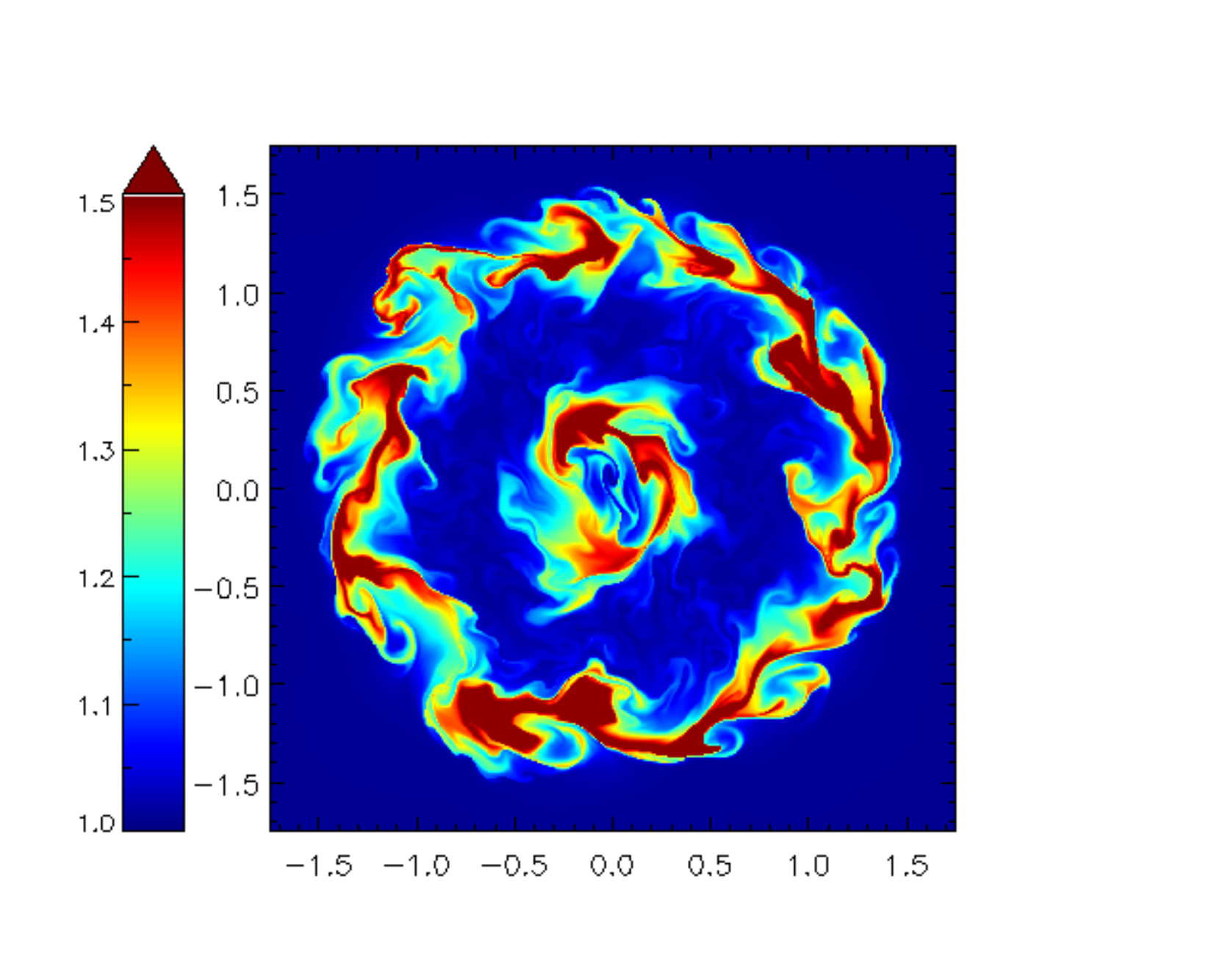}
\centering\includegraphics[scale=0.5,trim=0cm 0cm 3.5cm 0cm, clip=true]{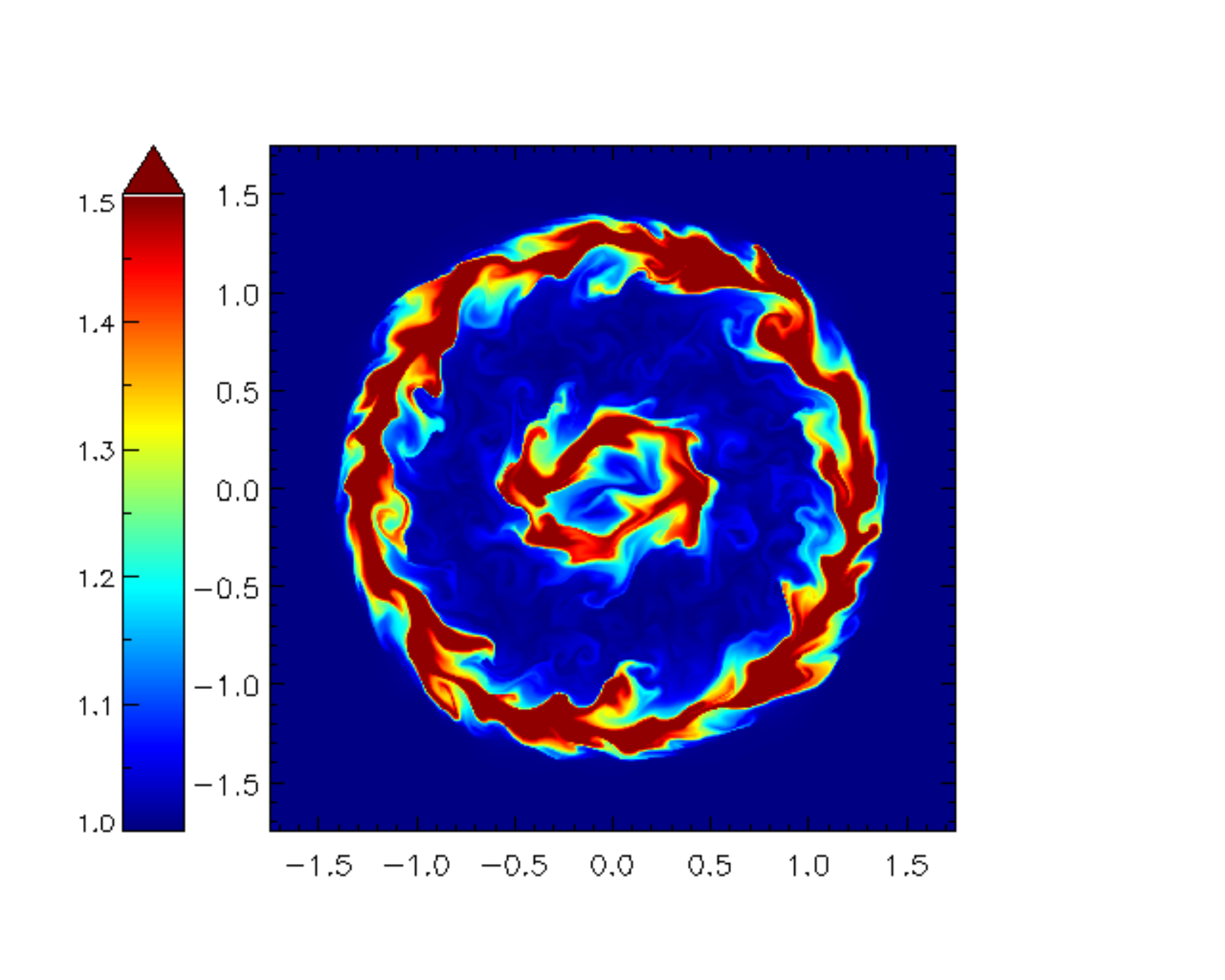}
\centering\includegraphics[scale=0.5,trim=0cm 0cm 3.5cm 0cm, clip=true]{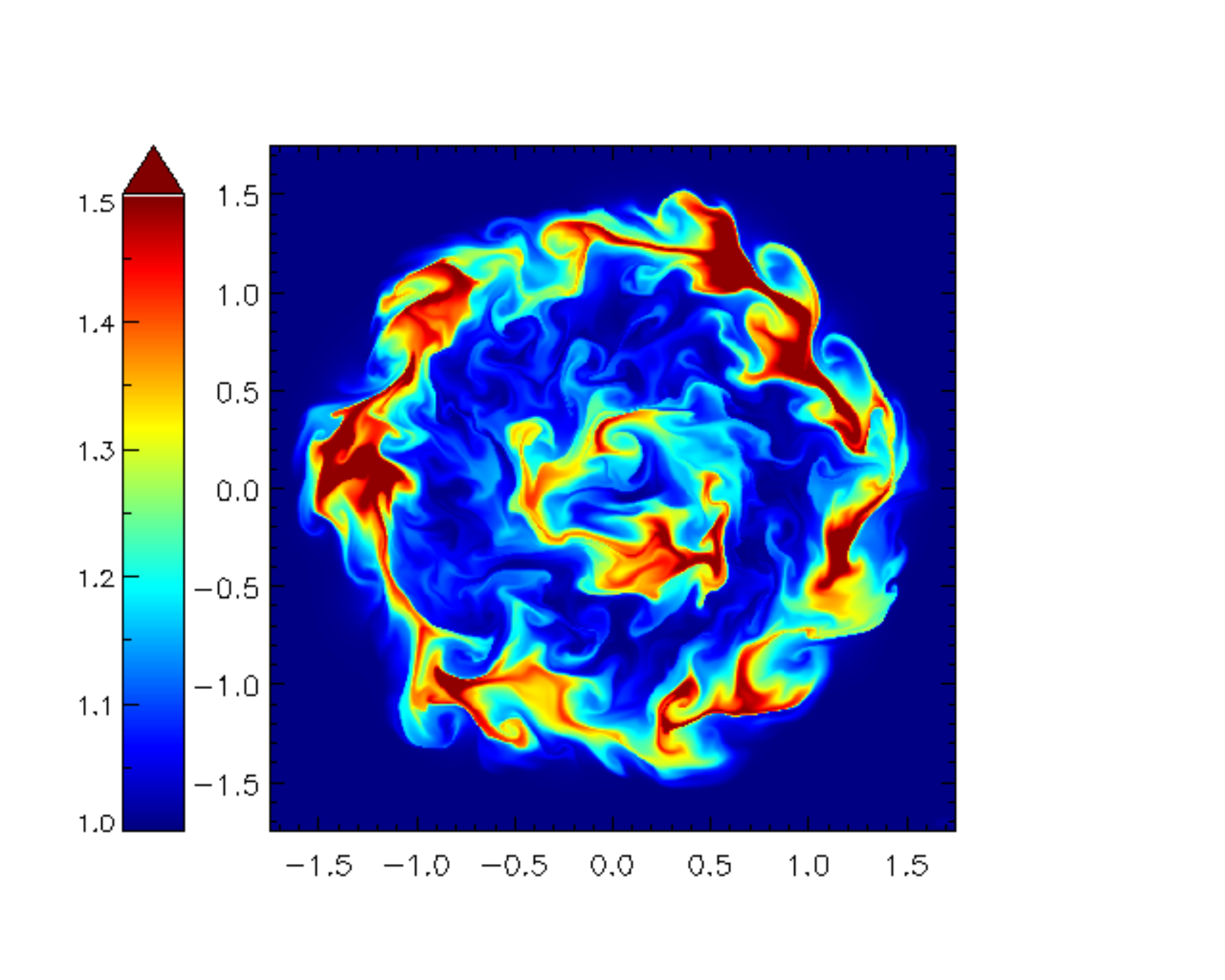}
\centering\includegraphics[scale=0.5,trim=0cm 0cm 3.5cm 0cm, clip=true]{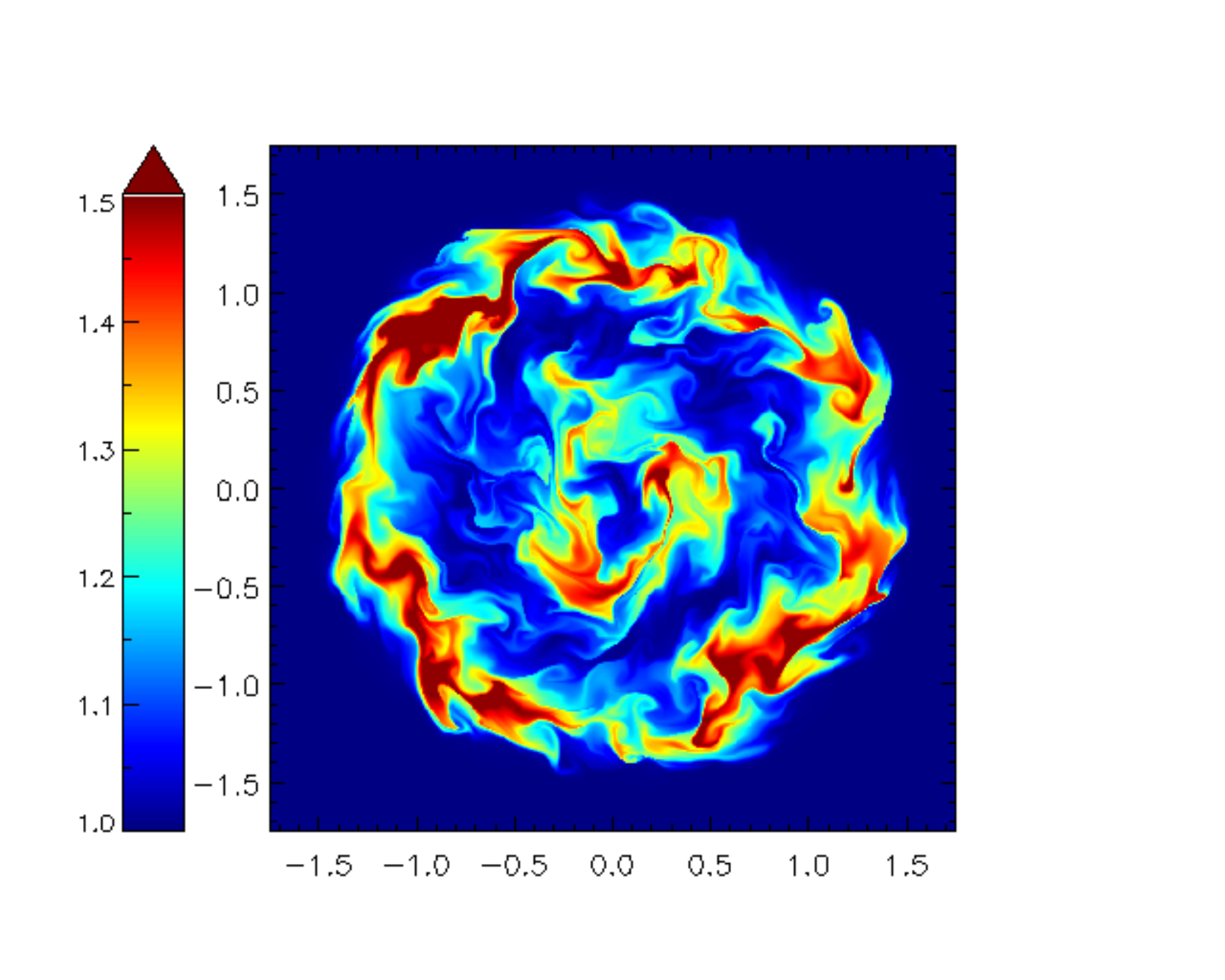}
\centering\includegraphics[scale=0.5,trim=0cm 0cm 3.5cm 0cm, clip=true]{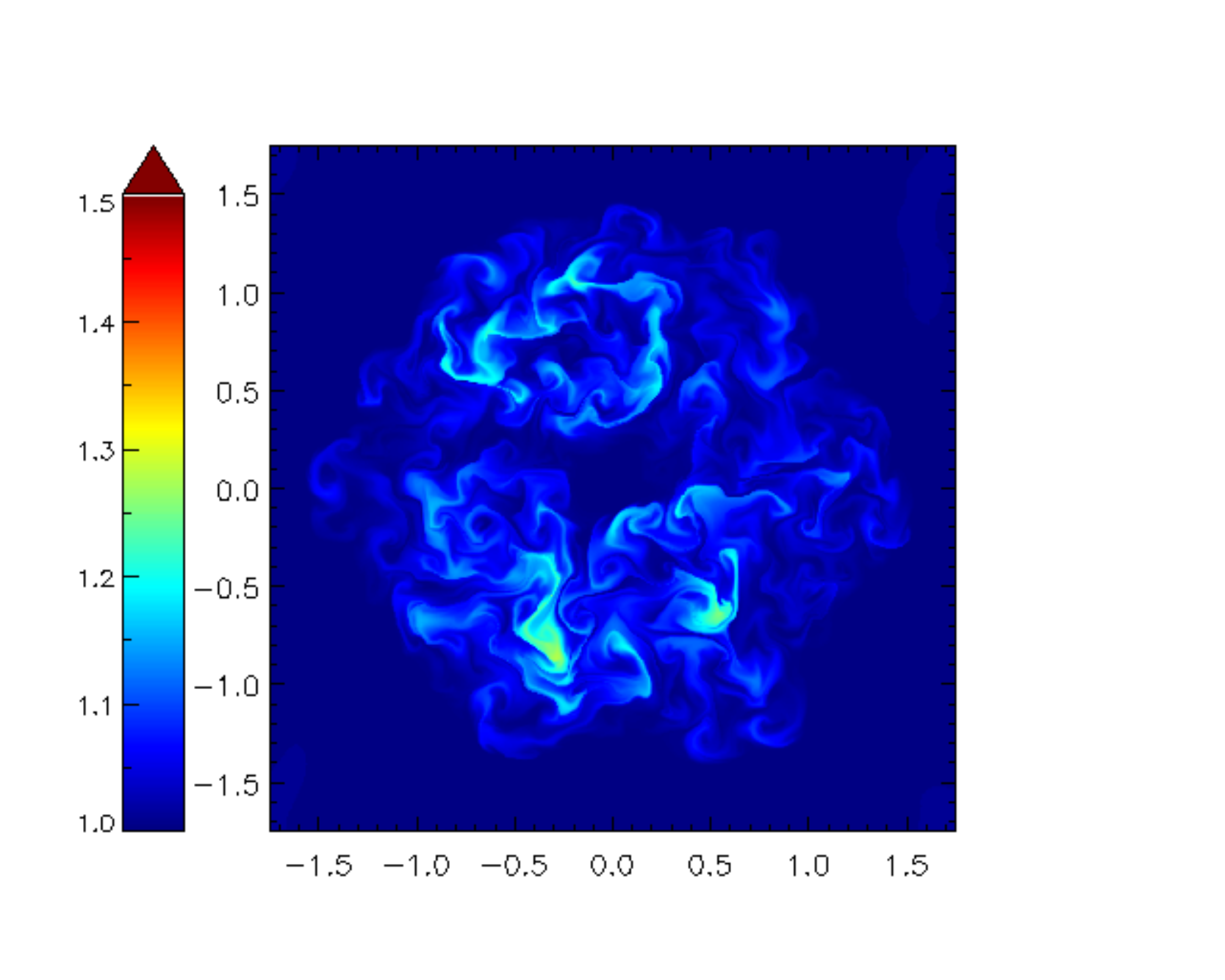}
\centering\includegraphics[scale=0.5,trim=0cm 0cm 3.5cm 0cm, clip=true]{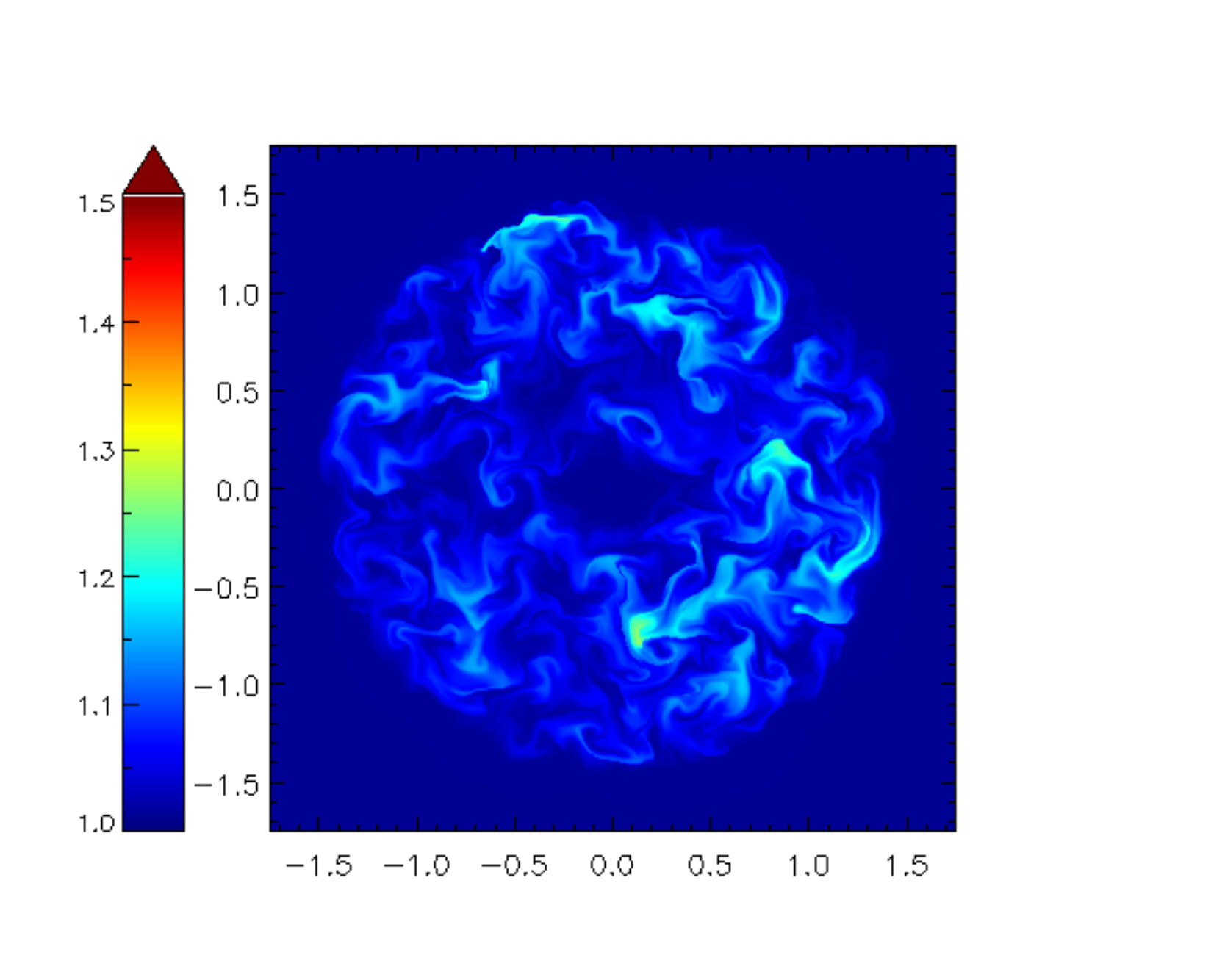}
\caption{Maps of field line length at the end of the fixed-pattern (left) and random-pattern (right) cases with $k=1$ (top row), $k=0.75$ (middle row), and $k=0.5$ (bottom row).}
\label{fig:Blength}
\end{figure*}

 Perhaps the clearest example of the effect of helicity preference on the structure of the closed loop corona is evident in Figures \ref{fig:structure_f} and \ref{fig:structure_r}. The smoothness of the $k=1$ and $k=0.75$ coronal loops is manifestly different than the complexity of the $k=0.5$ coronal loops. Observations of the coronal magnetic field, meanwhile, invariably reveal smooth, laminar loops that closely resemble those observed in the $k=1$ and $k=0.75$ cases (cf. Figure \ref{fig:obs}), rather than those observed in the $k=0.5$ case. Our simulations indicate, therefore, that the photosphere must inject a significant net helicity so that structures such as those seen in the $k=0.5$ are not observed.

The simulation setups assumed in this paper are  simplified
compared to the complex photospheric polarity structure exhibited by the Sun, illustrated by
Figure \ref{fig:obs}. Nevertheless, the quasi-random structure that we
obtained for a 50\% helicity preference, shown in the bottom panels of
Figures \ref{fig:Bphi_end} and \ref{fig:Blength}, obviously does not
 resemble the clean, bimodal structure observed for the
corona. In contrast, our results for both 100\% and 75\% preferences,
shown in the other panels of those figures, do exhibit the bimodal
characteristics of the corona: concentrations of twist at PILs in the
form of highly nonpotential magnetic shear in filament channels, and
generally smooth, quasi-potential fields free of twist away from PILs
in arcades of coronal loops. Therefore, a principal conclusion of our
work is that the Sun must inject helicity into the corona with a
significant hemispheric preference, favoring negative helicity forming
left-handed structures in the north, and positive helicity forming
right-handed structures in the south. These preferences are reflected
in the observed statistics of solar filaments, sigmoids, and sunspot
whorls. They also have been detected directly in the photospheric
convection, although that measurement is very challenging, near the
limits of observational resolution.\par

Our simulation is simplified in another important way compared to the
Sun: there is no source of new, weakly sheared or unsheared magnetic
flux in our domain, nor is there a sink of the strongly sheared flux
condensing in the filament channels. Flux emergence from below the
photosphere constantly injects fresh magnetic field into the corona,
and coronal mass ejections regularly eject sheared magnetic field and
its entrained magnetic helicity away from the Sun into the
heliosphere. The characteristic time scales for these phenomena
compete directly with the filament-channel formation time $\tau_{fc}$
to establish a quasi-steady balance among these processes and the
coronal magnetic structure that is observed. Such a calculation is
well beyond the scope of this paper, but a first attack on the problem
could be taken using global force-free modeling of the corona
\citep[e.g.\ ][]{Mackay14}. We point out that $\tau_{fc}$ in Equation
(\ref{Taufc}) is inversely proportional to the product of the net
fractional helicity $f$ and the angular rotation rate $\omega_0$ of
the twisting motions ($H_0 \propto \omega_0 \tau_0$; KAD15). If this
product $f \omega_0$ is too small or too large, then the
filament-channel formation time will be too long or too short compared
to the emergence and ejection time scales, and the model is unlikely
to replicate the Sun's observed appearance. We anticipate that global
modeling of the combined processes could provide rigorous bounds on
the rotation rate $\omega_0$, to complement the narrowly constrained
range of values available to the net fractional helicity,
$0.5 < f < 1.0$.\par

Our simulations show that random displacements of the pattern of
photospheric convection have only a secondary effect on the resulting
coronal structure. This also is evident in Figures \ref{fig:Bphi_end}
and \ref{fig:Blength} by comparing the left (fixed-pattern) and right
(random-pattern) columns for each helicity preference. The latter
structures are somewhat smoother than the former, especially at small
scales, but the large-scale organization is no different between
them. This conclusion agrees with that reached by \citet{Zhao15}, who
used a much simpler setup with far fewer rotation cells. The
fundamental reason that the random convection pattern shows a very
similar coronal end state to that of the fixed pattern is that, as shown by
numerous simulations, magnetic reconnection is efficient at
destroying all higher order topological features such as braiding,
leaving only the global helicity \citep[e.g.,][]{WilmotSmith10}. Our
results are fully in agreement with this hypothesis. \par

In summary, this paper presents the first simulations of the evolution
of the coronal magnetic field driven by photospheric motions with
varying helicity preference.  Our results agree well with the helicity
condensation model of \citet{Antiochos13}, which accounts for both the
formation of sheared filament channels adjacent to PILs and the
quasi-potential, smooth character of coronal loops away from PILs. By
transferring the magnetic twist injected by photospheric motions to
ever larger scales, reconnection concentrates the twist at the
boundaries of flux systems (i.e., at the PILs) while diluting it
throughout their interiors. The remarkable implication of the model is
that the global organization of the magnetic shear in the solar
atmosphere is a direct consequence of local twisting of the footpoints
of coronal flux tubes by surface convection. Even more remarkable and
somewhat counter-intuitive, is our finding that in order for the hot
closed corona -- the loops -- to exhibit no structure such as tangling
or twisting, then a great deal of structure needs to be injected!  The
photospheric driving motions must have a clear helicity preference, in
which case all the injected structure ends up localized around
PILs. The corona, therefore, is a striking example on cosmic scales of
a strongly self-organized system. \par

\acknowledgments{
K.J.K acknowledges the use of post-processing codes originally written by Benjamin Lynch and Peter Wyper. K.J.K received funding for this work through a NASA Earth and Space Science Fellowship. The numerical simulations were performed under a grant of High-End Computing resources to C.R.D. at NASA's Center for Climate Simulation. S.K.A. and C.R.D. were supported, in part, by grants from NASA's Living With a Star and Heliophysics Supporting Research programs.
}


\clearpage

\newpage

\begin{thebibliography}{}
\expandafter\ifx\csname natexlab\endcsname\relax\def\natexlab#1{#1}\fi

\bibitem[{{Antiochos}(2013)}]{Antiochos13}
{Antiochos}, S.~K. 2013, \apj, 772, 72

\bibitem[{{Attie} {et~al.}(2009){Attie}, {Innes}, \& {Potts}}]{Attie09}
{Attie}, R., {Innes}, D.~E., \& {Potts}, H.~E. 2009, \aap, 493, L13

\bibitem[{{Berger}(1984)}]{Berger84b}
{Berger}, M.~A. 1984, GApFD, 30, 79

\bibitem[{{Bonet} {et~al.}(2008){Bonet}, {M{\'a}rquez}, {S{\'a}nchez Almeida},
  {Cabello}, \& {Domingo}}]{Bonet08}
{Bonet}, J.~A., {M{\'a}rquez}, I., {S{\'a}nchez Almeida}, J., {Cabello}, I., \&
  {Domingo}, V. 2008, \apjl, 687, L131

\bibitem[{{Bonet} {et~al.}(2010){Bonet}, {M{\'a}rquez}, {S{\'a}nchez Almeida},
  {Palacios}, {Mart{\'{\i}}nez Pillet}, {Solanki}, {del Toro Iniesta},
  {Domingo}, {Berkefeld}, {Schmidt}, {Gandorfer}, {Barthol}, \&
  {Kn{\"o}lker}}]{Bonet10}
{Bonet}, J.~A., {M{\'a}rquez}, I., {S{\'a}nchez Almeida}, J., {et~al.} 2010,
  \apjl, 723, L139

\bibitem[{{Brandt} {et~al.}(1988){Brandt}, {Scharmer}, {Ferguson}, {Shine}, \&
  {Tarbell}}]{Brandt88}
{Brandt}, P.~N., {Scharmer}, G.~B., {Ferguson}, S., {Shine}, R.~A., \&
  {Tarbell}, T.~D. 1988, Natur, 335, 238

\bibitem[{DeVore(1991)}]{DeVore91}
DeVore, C.~R. 1991, JCoPh, 92, 142

\bibitem[{{DeVore} \& {Antiochos}(2008)}]{DeVore08}
{DeVore}, C.~R., \& {Antiochos}, S.~K. 2008, \apj, 680, 740

\bibitem[{{Duvall} \& {Gizon}(2000)}]{Duvall00}
{Duvall}, Jr., T.~L., \& {Gizon}, L. 2000, SoPh, 192, 177

\bibitem[{Finn \& Antonsen(1985)}]{Finn85}
Finn, J., \& Antonsen, T. 1985, CoPPC, 9, 111

\bibitem[{{Gaizauskas}(2000)}]{Gaizauskas00}
{Gaizauskas}, V. 2000, in Encyclopedia of Astronomy and Astrophysics, ed.
  P.~{Murdin} (Bristol: IOP), 2278

\bibitem[{{Gizon} \& {Duvall}(2003)}]{Gizon03}
{Gizon}, L., \& {Duvall}, Jr., T.~L. 2003, in ESA Special Publication, Vol.
  517, GONG+ 2002 Local and Global Helioseismology: the Present and Future, ed.
  H.~{Sawaya-Lacoste} (Noordwijk: ESA), 43

\bibitem[{{Hagino} \& {Sakurai}(2002)}]{Hagino02}
{Hagino}, M., \& {Sakurai}, T. 2002, in Multi-Wavelength Observations of
  Coronal Structure and Dynamics, ed. P.~C.~H. {Martens} \& D.~{Cauffman}
  (Pergamon: Amsterdam), 147

\bibitem[{{Hale}(1927)}]{Hale27}
{Hale}, G.~E. 1927, \nat, 119, 708

\bibitem[{{Hirzberger} {et~al.}(2008){Hirzberger}, {Gizon}, {Solanki}, \&
  {Duvall}}]{Hirzberger08}
{Hirzberger}, J., {Gizon}, L., {Solanki}, S.~K., \& {Duvall}, T.~L. 2008, SoPh,
  251, 417

\bibitem[{{Knizhnik} {et~al.}(2015){Knizhnik}, {Antiochos}, \&
  {DeVore}}]{Knizhnik15}
{Knizhnik}, K.~J., {Antiochos}, S.~K., \& {DeVore}, C.~R. 2015, \apj, 809, 137

\bibitem[{{Komm} {et~al.}(2007){Komm}, {Howe}, {Hill}, {Miesch}, {Haber}, \&
  {Hindman}}]{Komm07}
{Komm}, R., {Howe}, R., {Hill}, F., {et~al.} 2007, \apj, 667, 571

\bibitem[{{Liu} \& {Schuck}(2012)}]{Liu12}
{Liu}, Y., \& {Schuck}, P.~W. 2012, \apj, 761, 105

\bibitem[{{Mackay} {et~al.}(2014){Mackay}, {DeVore}, \& {Antiochos}}]{Mackay14}
{Mackay}, D.~H., {DeVore}, C.~R., \& {Antiochos}, S.~K. 2014, \apj, 784, 164

\bibitem[{{Martin}(1994)}]{Martin94}
{Martin}, S.~F. 1994, in ASP Conf. Ser., Vol.~68, Solar Active Region
  Evolution: Comparing Models with Observations, ed. K.~S. {Balasubramaniam} \&
  G.~W. {Simon} (San Francisco: ASP), 264

\bibitem[{{Martin}(1998)}]{Martin98}
{Martin}, S.~F. 1998, SoPh, 182, 107

\bibitem[{{Martin} {et~al.}(1992){Martin}, {Marquette}, \&
  {Bilimoria}}]{Martin92}
{Martin}, S.~F., {Marquette}, W.~H., \& {Bilimoria}, R. 1992, in ASP Conf.
  Ser., Vol.~27, The Solar Cycle, ed. K.~L. {Harvey} (San Francisco: ASP), 53

\bibitem[{{Parker}(1972)}]{Parker72}
{Parker}, E.~N. 1972, \apj, 174, 499

\bibitem[{{Pevtsov} {et~al.}(2003){Pevtsov}, {Balasubramaniam}, \&
  {Rogers}}]{Pevtsov03}
{Pevtsov}, A.~A., {Balasubramaniam}, K.~S., \& {Rogers}, J.~W. 2003, \apj, 595,
  500

\bibitem[{{Pevtsov} {et~al.}(2014){Pevtsov}, {Berger}, {Nindos}, {Norton}, \&
  {van Driel-Gesztelyi}}]{Pevtsov14}
{Pevtsov}, A.~A., {Berger}, M.~A., {Nindos}, A., {Norton}, A.~A., \& {van
  Driel-Gesztelyi}, L. 2014, SSRv, 186, 285

\bibitem[{{Pevtsov} \& {Longcope}(2001)}]{Pevtsov01b}
{Pevtsov}, A.~A., \& {Longcope}, D.~W. 2001, in ASP Conf. Ser., Vol. 236,
  Advanced Solar Polarimetry -- Theory, Observation, and Instrumentation, ed.
  M.~{Sigwarth} (San Francisco: ASP), 423

\bibitem[{{Rappazzo} {et~al.}(2013){Rappazzo}, {Velli}, \&
  {Einaudi}}]{Rappazzo13}
{Rappazzo}, A.~F., {Velli}, M., \& {Einaudi}, G. 2013, \apj, 771, 76

\bibitem[{{Rust} \& {Kumar}(1994)}]{Rust94b}
{Rust}, D.~M., \& {Kumar}, A. 1994, SoPh, 155, 69

\bibitem[{{Rust} \& {Kumar}(1996)}]{Rust96}
---. 1996, \apjl, 464, L199

\bibitem[{{Schmieder} {et~al.}(2014){Schmieder}, {Roudier}, {Mein}, {Mein},
  {Malherbe}, \& {Chandra}}]{Schmieder14}
{Schmieder}, B., {Roudier}, T., {Mein}, N., {et~al.} 2014, \aap, 564, A104

\bibitem[{{Schrijver} {et~al.}(1999){Schrijver}, {Title}, {Berger}, {Fletcher},
  {Hurlburt}, {Nightingale}, {Shine}, {Tarbell}, {Wolfson}, {Golub},
  {Bookbinder}, {Deluca}, {McMullen}, {Warren}, {Kankelborg}, {Handy}, \& {de
  Pontieu}}]{Schrijver99}
{Schrijver}, C.~J., {Title}, A.~M., {Berger}, T.~E., {et~al.} 1999, SoPh, 187,
  261

\bibitem[{{Seehafer}(1990)}]{Seehafer90}
{Seehafer}, N. 1990, SoPh, 125, 219

\bibitem[{{Seligman} {et~al.}(2014){Seligman}, {Petrie}, \&
  {Komm}}]{Seligman14}
{Seligman}, D., {Petrie}, G.~J.~D., \& {Komm}, R. 2014, \apj, 795, 113

\bibitem[{{Taylor}(1974)}]{Taylor74}
{Taylor}, J.~B. 1974, PhRvL, 33, 1139

\bibitem[{{Taylor}(1986)}]{Taylor86}
---. 1986, RvMP, 58, 741

\bibitem[{{Vargas Dom{\'{\i}}nguez} {et~al.}(2011){Vargas Dom{\'{\i}}nguez},
  {Palacios}, {Balmaceda}, {Cabello}, \& {Domingo}}]{VD11}
{Vargas Dom{\'{\i}}nguez}, S., {Palacios}, J., {Balmaceda}, L., {Cabello}, I.,
  \& {Domingo}, V. 2011, \mnras, 416, 148

\bibitem[{{Vargas Dom{\'{\i}}nguez} {et~al.}(2015){Vargas Dom{\'{\i}}nguez},
  {Palacios}, {Balmaceda}, {Cabello}, \& {Domingo}}]{VD15}
---. 2015, SoPh, 290, 301

\bibitem[{{Wilmot-Smith} {et~al.}(2010){Wilmot-Smith}, {Pontin}, \&
  {Hornig}}]{WilmotSmith10}
{Wilmot-Smith}, A.~L., {Pontin}, D.~I., \& {Hornig}, G. 2010, \aap, 516, A5

\bibitem[{{Woltjer}(1958)}]{Woltjer58}
{Woltjer}, L. 1958, PNAS, 44, 489

\bibitem[{{Zhao} {et~al.}(2015){Zhao}, {DeVore}, {Antiochos}, \&
  {Zurbuchen}}]{Zhao15}
{Zhao}, L., {DeVore}, C.~R., {Antiochos}, S.~K., \& {Zurbuchen}, T.~H. 2015,
  \apj, 805, 61

\bibitem[{{Zirker} {et~al.}(1997){Zirker}, {Martin}, {Harvey}, \&
  {Gaizauskas}}]{Zirker97}
{Zirker}, J.~B., {Martin}, S.~F., {Harvey}, K., \& {Gaizauskas}, V. 1997, SoPh,
  175, 27

\end{thebibliography}
\end{document}